\documentclass[iop,floatfix]{emulateapj}

\RequirePackage{color}
\RequirePackage{code}
\input{astro.sty}

\shorttitle{PS1 Calibration}
\shortauthors{E.A. Magnier et al}
\begin{document}
\title{Pan-STARRS Photometric and Astrometric Calibration}

\def\IfA{1}
\def\LBL{2}
\def\Hubble{3}
\def\ITC{4}
\def\Harvard{5}
\def\MPIA{6}
\def\ARI{7}
\def\Princeton{8}
\def\DUR{9}
\def\CfA{10}

\author{
Eugene. A. Magnier,\altaffilmark{\IfA}
Edward. F. Schlafly,\altaffilmark{\LBL,\Hubble}
Douglas P. Finkbeiner,\altaffilmark{\ITC,\Harvard}
J.~L. Tonry,\altaffilmark{\IfA}
B. Goldman,\altaffilmark{\MPIA}
S. R\"oser,\altaffilmark{\ARI}
E. Schilbach,\altaffilmark{\ARI}
K.~C. Chambers,\altaffilmark{\IfA} 
H.~A. Flewelling,\altaffilmark{\IfA}
M. E. Huber,\altaffilmark{\IfA}
P.~A. Price,\altaffilmark{\Princeton}
W.~E. Sweeney,\altaffilmark{\IfA}
C. Z. Waters,\altaffilmark{\IfA}
L. Denneau,\altaffilmark{\IfA}
P. Draper,\altaffilmark{\DUR}
K. W. Hodapp,\altaffilmark{\IfA}
R. Jedicke,\altaffilmark{\IfA}
N. Kaiser,\altaffilmark{\IfA}
R.-P. Kudritzki,\altaffilmark{\IfA}
N. Metcalfe,\altaffilmark{\DUR}
C.~W. Stubbs,\altaffilmark{\CfA}
R. J. Wainscoat\altaffilmark{\IfA}
} 

\altaffiltext{\IfA}{Institute for Astronomy, University of Hawaii, 2680 Woodlawn Drive, Honolulu HI 96822}
\altaffiltext{\LBL}{Lawrence Berkeley National Laboratory, One Cyclotron Road, Berkeley, CA 94720, USA}
\altaffiltext{\Hubble}{Hubble Fellow}
\altaffiltext{\ITC}{Institute for Theory and Computation, Harvard-Smithsonian Center for Astrophysics, 60 Garden Street, MS-51, Cambridge, MA 02138 USA}
\altaffiltext{\Harvard}{Department of Physics, Harvard University, Cambridge, MA 02138 USA}
\altaffiltext{\MPIA}{Max Planck Institute for Astronomy, K\"onigstuhl 17, D-69117 Heidelberg, Germany}
\altaffiltext{\ARI}{Astronomisches Rechen-Institut, Zentrum f\"ur Astronomie der Universit\"at Heidelberg, M\"ochhofstrasse 12-14, D-69120 Heidelberg, Germany}
\altaffiltext{\Princeton}{Department of Astrophysical Sciences, Princeton University, Princeton, NJ 08544, USA}
\altaffiltext{\DUR}{Department of Physics, Durham University, South Road, Durham DH1 3LE, UK}
\altaffiltext{\CfA}{Harvard-Smithsonian Center for Astrophysics, 60 Garden Street, Cambridge, MA 02138}



\begin{abstract}

We present the details of the photometric and astrometric calibration
of the Pan-STARRS\,1 $3\pi$ Survey.  The photometric goals were to
reduce the systematic effects introduced by the camera and detectors,
and to place all of the observations onto a photometric system with
consistent zero points over the entire area surveyed, the \approx
30,000 square degrees north of $\delta = -30$\degrees.  The
astrometric calibration compensates for similar systematic effects so
that positions, proper motions, and parallaxes are reliable as well.
The Pan-STARRS Data Release 2 (DR2) astrometry is tied to the Gaia DR1
release.

\end{abstract}


\section{Introduction}\label{sec:intro}

From May 2010 through March 2014, the Pan-STARRS Science Consortium
used the 1.8m \PSONE\ telescope to perform a set of wide-field science
surveys.  These surveys are designed to address a range of science
goals included the search for hazardous asteroids, the study of the
formation and architecture of the Milky Way galaxy, and the search for
Type Ia supernovae to measure the history of the expansion of the
universe.  The majority of the time (56\%) was spent on surveying the
$\frac{3}{4}$ of the sky north of $-30$ Declination with
\grizy\ filters in the so-called $3\pi$ Survey.  Another $\sim 25\%$
of the time was concentrated on repeated deep observations of 10
specific fields in the Medium-Deep Survey.  The rest of the time was
used for several other surveys, including a search for potentially
hazardous asteroids in our solar system.  The details of the
telescope, surveys, and resulting science publications are described
by \cite{chambers2017}.

The wide-field \PSONE\ telescope consists of a 1.8~meter diameter
$f$/4.4 primary mirror with an 0.9~m secondary, producing a 3.3 degree
field of view \citep{2004SPIE.5489..667H}.  The optical design yields
low distortion and minimal vignetting even at the edges of the
illuminated region.  The optics, in combination with the natural
seeing, result in generally good image quality: the median image
quality for the 3$\pi$ survey is FWHM = (1.31, 1.19, 1.11, 1.07, 1.02)
arcseconds for (\grizy), with a floor of $\sim0.7$ arcseconds.  The
\PSONE\ camera \citep{2009amos.confE..40T} is a mosaic of 60
edge-abutted $4800\times4800$ pixel back-illuminated CCID58 Orthogonal
Transfer Arrays manufactured by Lincoln Laboratory
\citep{2006amos.confE..47T,2008SPIE.7021E..05T}.  The CCDs have
10~$\mu$m pixels subtending 0.258~arcsec and are 70$\mu$m thick.  The
detectors are read out using a StarGrasp CCD controller, with a
readout time of 7 seconds for a full unbinned image
\citep{2008SPIE.7014E..0DO}.  The active, usable pixels cover $\sim
80$\% of the FOV.

Nightly observations are conducted remotely from the Advanced
Technology Research Center in Kula, the main facility of the
University of Hawaii's Institute for Astronomy operations on Maui.
During the \PSONE\ Science Survey, images obtained by the
\PSONE\ system were stored first on computers at the summit, then
copied with low latency via internet to the dedicated data analysis
cluster located at the Maui High Performance Computer Center in Kihei,
Maui.

Pan-STARRS produced its first large-scale public data release, Data
Release 1 (DR1) on 16 December 2016.  DR1 contains the results of the
third full reduction of the Pan-STARRS $3\pi$ Survey archival data,
identified as PV3.  Previous reductions \citep[PV0, PV1, PV2;
 see][]{magnier2017.datasystem} were used internally for pipeline
optimization and the development of the initial photometric and
astrometric reference catalog \citep{magnier2017.calibration}.  The
products from these reductions were not publicly released, but have
been used to produce a wide range of scientific papers from the
Pan-STARRS 1 Science Consortium members \citep{chambers2017}.  DR1
contained only average information resulting from the many individual
images obtained by the $3\pi$ Survey observations.  A second data
release, DR2, was made available 28 January 2019.  DR2 provides
measurements from all of the individual exposures, and include an
improved calibration of the PV3 processing of that dataset.

This is the fifth in a series of seven papers describing the
Pan-STARRS1 Surveys, the data reduction techniques and the resulting
data products.  This paper (Paper V) describes the final calibration
process, and the resulting photometric and astrometric quality.

\citet[][Paper I]{chambers2017}
provides an overview of the Pan-STARRS System, the design and
execution of the Surveys, the resulting image and catalog data
products, a discussion of the overall data quality and basic
characteristics, and a brief summary of important results.

\citet[][Paper II]{magnier2017.datasystem}
describes how the various data processing stages are organized and implemented
in the Imaging Processing Pipeline (IPP), including details of the 
the processing database which is a critical element in the IPP infrastructure . 

\citet[][Paper III]{waters2017} describes the details of the pixel
processing algorithms, including detrending, warping, and adding (to
create stacked images) and subtracting (to create difference images)
and resulting image products and their properties.

\citet[][Paper IV]{magnier2017.analysis} describes the details of the source
detection and photometry, including point-spread-function and extended
source fitting models, and the techniques for ``forced" photometry
measurements.


\citet[][Paper VI]{flewelling2017}
describes  the details of the resulting catalog data and its organization in the Pan-STARRS database. 

\citet[][Paper VII]{huber2017} describes the Medium Deep Survey in
detail, including the unique issues and data products specific to that
survey. The Medium Deep Survey is not part of Data Releases 1 or 2 and
will be made available in a future data release.

The Pan-STARRS1 filters and photometric system have already been
described in detail in \cite{2012ApJ...750...99T}.


\section{Pan-STARRS\,1 Data Analysis} 

Images obtained by \PSONE\ are automatically processed in real time by
the \PSONE\ Image Processing Pipeline \citep[IPP,][]{magnier2017.datasystem}.
Real-time analysis goals are aimed at feeding the discovery pipelines
of the asteroid search and supernova search teams.  The data obtained
for the \PSONE\ Science Survey has also been used in three additional
complete re-processing of the data: Processing Versions 1, 2, and 3
(PV1, PV2, and PV3).  The real-time processing of the data is
considered ``PV0''.  Except as otherwise noted, this article describes
the calibration of the PV3 analysis of the data.  Between the first
(DR1) and second (DR2) data releases, improvements were made to the
calibration of both the photometry and astrometry, as described in
this article.

The data processing steps are described in detail by \cite{waters2017}
and \cite{magnier2017.datasystem,magnier2017.analysis}.  In summary, individual images
are detrended: non-linearity and bias corrections are applied, a dark
current model is subtracted and flat-field corrections are applied.
The \yps-band images are also corrected for fringing: a master fringe
pattern is scaled to match the observed fringing and subtracted.  Mask
and variance image arrays are generated with the detrend analysis and
carried forward at each stage of the IPP processing.  Source detection
and photometry are performed for each chip independently.  As
discussed below, preliminary astrometric and photometric calibrations
are performed for all chips in a single exposure in a single analysis.
We refer to these measurements as the ``chip'' photometry and
astrometry products.

Chip images are geometrically transformed based on the astrometric
solution into a set of pre-defined pixel grids covering the sky,
called skycells.  These transformed images are called the warp images.
Sets of warps for a given part of the sky and in the same filter may
be added together to generate deeper `stack' images.  PSF-matched
difference images are generated from combinations of warps and stacks;
the details of the difference images and their calibration are outside
of the scope of this article.


Astronomical objects are detected and characterized in the stack
images.  The details of the analysis of the sources in the stack
images are discussed in \cite{magnier2017.analysis}, but in brief
these include PSF photometry, along with a range of measurements
driven by the goals of understanding the galaxies in the images.
Because of the significant mask fraction of the GPC1 focal plane, and
the varying image quality both within and between exposures, the
effective PSF of the PS1 stack images (often including more than 10
input exposures taken in different conditions) is highly variable.
The PSF varies significantly on scales as small as a few to tens of
pixels, making accurate PSF modelling essentially infeasible.  The PSF
photometry of sources in the stack images is thus degraded
significantly compared to the quality of the photometry measured for
the individual chip images.

To recover most of the photometric quality of the individual chip
images, while also exploiting the depth afforded by the stacks, the
PV3 analysis makes use of forced photometry on the individual warp
images.  PSF photometry is measured on the warp images for all sources
which are detected in the stack images images.  The positions
determined in the stack images are used in the warp images, but the
PSF model is determined for each warp independently based on brighter
stars in the warp image.  The only free parameter for each object is
the flux, which may be insignificant or even negative for sources
which are near the faint limit of the stack detections.  When the
fluxes from the individual warp images are averaged, a reliable
measurement of the faint source flux is determined.  The details of
this analysis are described in detail in \cite{magnier2017.analysis}.

The data products from the chip photometry, stack photometry, and
forced-warp photometry analysis stages are ingested into the internal
calibration database called the Desktop Virtual Observatory, or DVO
\citep[see Section~4 in][]{magnier2017.datasystem} and used for
photometric and astrometric calibrations.  In this article, we discuss
the photometric calibration of the individual exposures, the stacks,
and the warp images.  We also discuss the astrometric calibration of
the individual exposures and the stack images.

\section{Astrometric Models} 


Three somewhat distinct astrometric models are employed within the IPP
at different stages.  The simplest model is defined independently for
each chip: a simple TAN projection as described by
\cite{2002AA...395.1077C} is used to relate sky coordinates to a
Cartesian tangent-plane coordinate system.  A pair of low-order
polynomials are used to relate the chip pixel coordinates to this
tangent-plane coordinate system.  The transforming polynomials are of
the form:
\begin{eqnarray}
P & = & \sum_{i,j} C^P_{i,j} X^i_{\rm chip} Y^j_{\rm chip} \\
Q & = & \sum_{i,j} C^Q_{i,j} X^i_{\rm chip} Y^j_{\rm chip}
\end{eqnarray}
where $P,Q$ are the tangent plane coordinates, $X_{\rm chip}, Y_{\rm
  chip}$ are the coordinates on the 60 GPC1 chips, and $C^P_{i,j}, C^Q_{i,j}$
are the polynomial coefficients for each order.  In the \ippprog{psastro}
analysis, $i + j <= N_{\rm order}$ where the order of the fit, $N_{\rm
  order}$, may be 1 to 3, under the restriction that sufficient stars
are needed to constrain the order.  


A second form of astrometry model which yields somewhat higher
accuracy consists of a set of connected solutions for all chips in a
single exposure.  This model also uses a TAN projection to relate the
sky coordinates to a locally Cartesian tangent plane coordinate system.
A set of polynomials is then used to relate the tangent plane
coordinates to a `focal plane' coordinate system, $L,M$:
\begin{eqnarray}
P & = & \sum_{i,j} C^P_{i,j} L^i M^j \\
Q & = & \sum_{i,j} C^Q_{i,j} L^i M^j
\end{eqnarray}
This set of polynomials accounts for effects such as optical distortion
in the camera and distortions due to changing atmospheric refraction
across the field of the camera.  Since these effects are smooth across
the field of the camera, a single pair of polynomials can be used for
each exposure.  Like in the chip analysis above, the \ippprog{psastro}
code restricts the exponents with the rule $i + j <= N_{\rm order}$
where the order of the fit, $N_{\rm order}$, may be 1 to 3, under the
restriction that sufficient stars are needed to constrain the order
For each chip, a second set of polynomials describes the
transformation from the chip coordinate systems to the focal
coordinate system:
\begin{eqnarray}
L & = & \sum_{i,j} C^L_{i,j} X^i_{\rm chip} Y^j_{\rm chip} \\
M & = & \sum_{i,j} C^M_{i,j} X^i_{\rm chip} Y^j_{\rm chip}
\end{eqnarray}

A third form of the astrometry model is used in the context of the
calibration determined within the DVO database system.  We retain the
two levels of transformations (chip $\rightarrow$ focal plane $\rightarrow$
tangent plane), but the relationship between the chip and focal plane
is represented with only the linear terms in the polynomial,
supplemented by a coarse grid of displacements, $\delta L, \delta M$ sampled
across the coordinate range
of the chip.  This displacement grid may have a resolution of up to
$6\times6$ samples across the chip.  The displacement for a specific
chip coordinate value is determined via bilinear interpolation between
the nearest sample points.  Thus, the chip to focal-plane
transformation may be written as:
\begin{eqnarray}
  L & = & C^L_{0,0} + C^L_{1,0} X_{\rm chip} + C^L_{0,1} Y_{\rm chip} + \delta L(X_{\rm chip}, Y_{\rm chip}) \\
  M & = & C^M_{0,0} + C^M_{1,0} X_{\rm chip} + C^M_{0,1} Y_{\rm chip} + \delta M(X_{\rm chip}, Y_{\rm chip}) 
\end{eqnarray}





\section{Real-time Calibration}

\subsection{Overview}

As images are processed by the data analysis system, every exposure is
calibrated individually with respect to a photometric and astrometric
database.  The goal of this calibration step is to generate a preliminary
astrometric calibration, to be used by the warping analysis to determine
the geometric transformation of the pixels, and preliminary
photometric transformation, to be used by the stacking analysis to
ensure the warps are combined using consistent flux units.

The program used for the real-time calibration, \ippprog{psastro},
loads the measurements of the chip detections from their individual
output catalog files.  It uses the header information populated at the
telescope to determine an initial astrometric calibration guess based
on the position of the telescope boresite right ascension, declination
and position angle as reported by the telescope \& camera subsystems.
Using the initial guess, \ippprog{psastro} loads astrometric and
photometric data from the reference database.

\subsection{Reference Catalogs}
\label{sec:synthdb}

During the course of the PS1SC Survey, several reference databases
have been used.  For the first 20 months of the survey,
\ippprog{psastro} used a reference catalog with synthetic PS1
\grizy\ photometry generated by the Pan-STARRS IPP team based on based
combined photometry from Tycho (B, V), USNO \citep[red, blue,
  IR][]{2003AJ....125..984M}, and 2MASS
$J, H, K$ \citep{2006AJ....131.1163S}.  The astrometry in the database was from 2MASS
\citep{2006AJ....131.1163S}.  After 2012 May, a reference catalog
generated from internal re-calibration of the PV0 analysis of PS1
photometry and astrometry was used for the reference catalog.


Coordinates and calibrated magnitudes of stars from the reference
database are loaded by \code{pasastro}.  A model for the positions of
the 60 chips in the focal plane is used to determine the expected
astrometry for each chip based on the boresite coordinates and
position angle reported by the header.  Reference stars are selected
from the full field of view of the GPC1 camera, padded by an
additional 25\% to ensure a match can be determined even in the
presence of substantial errors in the boresite coordinates.  It is
important to choose an appropriate set of reference stars: if too few
are selected, the chance of finding a match between the reference and
observed stars is diminished.  In addition, since stars are loaded in
brightness order, a selection which is too small is likely to contain
only stars which are saturated in the GPC1 images.  On the other hand,
if too many reference stars are chosen, there is a higher chance of a
false-positive match, especially as many of the reference stars may
not be detected in the GPC1 image.  The selection of the reference
stars includes a limit on the brightest and faintest magnitudes of the
stars selected.

The astrometric analysis is necessarily performed first; after the
astrometry is determined, an automatic byproduct is a reliable match
between reference and observed stars, allowing a comparison of the
magnitudes to determine the photometric calibration.  

The astrometric calibration is performed in two major stages: first,
the chips are fitted independently with independent models for each
chip.  This fit is sufficient to ensure a reliable match between
reference stars and observed sources in the image.  Next, the set of
chip calibrations are used to define the transformation between the
focal plane coordinate system and the tangent plane coordinate
system.  The chip-to-focal plane transformations are then determined
under the single common focal plane to tangent plane transformation.  

\subsection{Cross-Correlation Search}

The first step of the analysis is to attempt to find the match between
the reference stars and the detected objects.  \ippprog{psastro} uses 2D
cross correlation to search for the match.  The guess astrometry
calibration is used to define a predicted set of $X^{\rm ref}_{\rm
  chip}, Y^{\rm ref}_{\rm chip}$ values for the reference catalog
stars.  For all possible pairs between the two lists, the values of
\begin{eqnarray}
\Delta X & = & X^{\rm ref}_{\rm chip} - X^{\rm obs}_{\rm chip}\\
\Delta Y & = & Y^{\rm ref}_{\rm chip} - Y^{\rm obs}_{\rm chip}
\end{eqnarray}
are generated.  The collection of $\Delta X, \Delta Y$ values are
collected in a 2D histogram with sampling of 50 pixels and the
peak pixel is identified.  If the astrometry guess were perfect, this
peak pixel would be expected to lie at (0,0) and contain all of the
matched stars.  However, the astrometric guess may be wrong in
several ways.  An error in the constant term above, $C^P_{0,0},
C^Q_{0,0}$ shifts the peak to another pixel, from which $C^P_{0,0},
C^Q_{0,0}$ can easily be determined.  An error in the plate scale or a
rotation will smear out the peak pixel potentially across many pixels
in the 2D histogram.  

To find a good match in the face of plate scale and rotation errors,
the cross correlation analysis above is performed for a series of
trials in which the scale and rotation are perturbed from the nominal
value by a small amount.  For each trial, the peak pixel is found and
a figure of merit is measured.  The figure of merit is defined as
$\frac{\sigma^2_x + \sigma^2_y}{N_p^4}$ where $\sigma^2_{x,y}$ is the
second moment of $\Delta X,Y$ for the star pairs associated with the
peak pixel, and $N_p$ is the number of star pairs in the peak.  This
figure of merit is thus most sensitive to a narrow distribution with
many matched pairs.  For the PS1 exposures, rotation offsets of (-1.0,
-0.5, 0.0, 0.5, 1.0) degrees, and plate scales of (+1\%, 0, -1\%) of
the nominal plate scale are tested.  The best match among these 15
cross-correlation tests is selected and used to generate a better
astrometry guess for the chip.


\subsection{Chip Polynomial Fits}

The astrometry solution from the cross correlation step above is again
used to select matches between the reference stars and observed
stars in the image.  The matching radius starts off quite large, and a
series of fits is performed to generate the transformation between
chip and tangent plane coordinates.  Three clipping iterations are
performed, with outliers $> 3 \sigma$ rejected on each pass, where
here $\sigma$ is determined from the distribution of the residuals in
each dimension (X,Y) independently.  After each fit cycle, the matches
are redetermined using a smaller radius and the fit re-tried.  

\subsection{Mosaic Astrometry Polynomial Fits}

The astrometry solutions from the independent chip fits are used to
generate a single model for the camera-wide distortion terms.  The
goal is to determine the two stage fit (chip $\rightarrow$ focal plane
$\rightarrow$ tangent plane).  There are a number of degenerate terms
between these two levels of transformation, most obviously between the
parameters which define the constant offset from chip to focal plane
($C^{L,M}_{0,0}$) and those which define the offset from focal plane
to tangent plane ($C^{P,Q}_{0,0}$).  We limit ($C^{P,Q}_{0,0}$) to be
0,0 to remove this degeneracy.  

The initial fit of the astrometry for each chip follows the distortion
introduced by the camera: the apparent plate scale for each chip is
the combination of the plate scale at the optical axis of the camera,
modified by the local average distortion.  To isolate the effect of
distortion, we choose a single common plate scale for the set of chips
and re-define the chip $\rightarrow$ sky calibrations as a set of chip
$\rightarrow$ focal plane transformations using that common pixel
scale.  We can now compare the observed focal plane coordinates,
derived from the chip coordinates, and the tangent plane coordinates,
derived from the projection of the reference coordinates.  One caveat
is that the chip reference coordinates are also degenerate with the
fitted distortion.  In order to avoid being sensitive to the exact
positions of the chips at this stage, we measure the local gradient
between the focal plane and tangent plane coordinate systems.  We then
fit the gradient with a polynomial of order 1 less than the polynomial
desired for the distortion fit.  The coefficients of the gradient fit
are then used to determine the coefficients for the polynomials
representing the distortion.  


Once the common distortion coming from the optics and atmosphere have
been modeled, \ippprog{psastro} determines polynomial transformations
from the 60 chips to the focal plane coordinate system.  In this
stage, 5 iterations of the chip fits are performed.  Before each
iteration, the reference stars and detected objects are matched using
the current best set of transformations.  These fits start with low
order (1) and large matching radius.  As the iterations proceed, the
radius is reduced and the order is allowed to increase, up to 3rd
order for the final iterations.  


\subsection{Real-time Photometric Calibration}


After the astrometric calibration has finished, the photometric
calibration is performed by \ippprog{psastro}.  When the reference
stars are loaded, the apparent magnitude in the filter of interest is
also loaded.  Stars for which the reference magnitude is brighter than
(\grizy) = (19, 19, 18.5, 18.5, 17.5) are used to determine the zero
points by comparison with the instrumental magnitudes.  For the PV3
analysis, an outlier-rejecting median is used to measure the zero
point. For early versions of the real-time analysis, when the
reference catalog used synthetic magnitudes, it was necessary to
search for the blue edge of the distribution: the synthetic magnitude
poorly predicted the magnitudes of stars in the presence of
significant extinction or for the very red stars, making the blue edge
somewhat more reliable as a reference than the mean.  Once the
calibration was based on a reference catalog generated from
\PSONE\ photometry, this methods was no longer needed.  Note that we
do not fit for the airmass slope in this analysis.  The nominal
airmass slope is used for each filter; any deviation from the nominal
value is effectively folded into the observed zero point.  The zero
point may be measured separately for each chip or as a single value
for the entire exposure; the latter option was used for the PV3
analysis.

\subsection{Real-time outputs}

The calibrations determined by \ippprog{psastro} are saved as part of
the header information in the output FITS tables.  For each exposure,
a single multi-extension FITS table is written.  In these files, the
measurements from each chip are written as a separate FITS table.  A
second FITS extension for each chip is used to store the header
information from the original chip image.  The original chip header is
modified so that the extension corresponds to an image with no pixel
data: \code{NAXIS} is set to 0, even though \code{NAXIS1} and
\code{NAXIS2} are retained with the original dimensions of the chip.
A pixel-less primary header unit (PHU) is generated with a summary of
some of the important and common chip-level keywords (e.g.,
\code{DATE-OBS}).  The astrometric transformation information for each
chip is saved in the corresponding header using standard (and some
non-standard) WCS keywords.  For the two-level astrometric model, the
PHU header carries the astrometric transformation related to the
projection and the camera-wide distortions.  Photometric calibrations
are written as a set of keywords to individual chip headers, and if
the calibration is performed at the exposure-level, to the PHU.  The
photometry calibration keywords are:
\begin{itemize}
\item \code{ZPT_REF} : the nominal zero point for this filter
\item \code{ZPT_OBS} : the measured zero point for this chip /
  exposure
\item \code{ZPT_ERR} : the measured error on \code{ZPT_OBS}
\item \code{ZPT_NREF} : the number of stars used to measure \code{ZPT_OBS}
\item \code{ZPT_MIN} : minimum reference magnitude included in analysis
\item \code{ZPT_MAX} : maximum reference magnitude included in analysis
\end{itemize}
The keyword \code{ZPT_OBS} is used to set the initial zero point when
the data from the exposure are loaded into the DVO database.

\section{PV3 DVO Master Database}

Data from the GPC1 chip images, the stack images, and the warp images
are loaded into the DVO calibration database using the real-time
analysis astrometric calibration to guide the association of
detections into objects.  After the full PV3 DVO database was
constructed, including all of the chip, stack, and warp detections,
several external catalogs were merged into the database.  First, the
complete 2MASS PSC was loaded into a stand-alone DVO database, which
was then merged into the PV3 master database.  Next the DVO database
of synthetic photometry in the PS1 bands (see
Section~\ref{sec:synthdb}) was merged in.  Next, the full Tycho
database was added, followed by the AllWISE database.  After the Gaia
release in August 2016 \citep{2016AA...595A...2G}, we generated a DVO
database of the Gaia positional and photometric information and merged
that into the master PV3 $3\pi$ DVO database.

The master DVO database is used to perform the full photometric and
astrometric calibration of the data.  During these analysis steps, a
wide variety of conditions are noted for individual measurements, for
the objects (either as a whole or for specific filters) and for the
images.  A set of bit-valued flags are used in the database to record
these conditions.
Table~\ref{tab:measure_mask_values} lists the flags specific to
individual measurements.  These values are stored in the DVO database in the
field \code{Measure.dbFlags} and exposed in the public database \citep[PSPS][]{flewelling2017}
in the fields \code{Detection.infoFlag3},
\code{StackObjectThin.XinfoFlag3} (where \code{X} is one of
     {$grizy$}), and \code{ForcedWarpMeasurement.FinfoFlag3}.
Table~\ref{tab:secf_mask_values} lists the flags which are set for
each filter for individual objects in the database.  These values are
recorded in the DVO database field \code{SecFilt.flags} and are
exposed in PSPS in the fields
\code{MeanObject.XFlags} and \code{StackObjectThin.XinfoFlag4}, where
\code{X} in both cases is one of {$grizy$}.
Table~\ref{tab:object_mask_values} lists the flags specific to an
object as a whole.  These values are stored in the DVO database field
\code{Average.flags} and are exposed in PSPS in
the field \code{MeanObject.objInfoFlag}.
Table~\ref{tab:image_mask_values} lists the flags raised for images.
These flags are stored in the DVO database field \code{Image.flags}
and are exposed in PSPS in the field \code{ImageMeta.qaFlags}.
The type of conditions which are recorded by these bits range from
information about the presence of external measurements (e.g., 2MASS
or WISE) to determinations of good or bad quality measurements for
astrometry or photometry.  In the sections below, these flag values in
these tables are described where appropriate.  Note that some of the
listed bits are either ephemeral (used internal to specific programs)
or are not relevant to the current DR2 analysis and reserved for
future use.

\begin{table*}
\begin{center}
\footnotesize
\caption{\label{tab:measure_mask_values} Per-Measurement Flag Bit Values} 
\begin{tabular}{lcl}
\hline
\hline
{\bf Bit Name} & {\bf Bit Value} & {\bf Description} \\
\hline
ID\_MEAS\_NOCAL              & 0x00000001 & detection ignored for this analysis (photcode, time range) -- internal only \\
ID\_MEAS\_POOR\_PHOTOM       & 0x00000002 & detection is photometry outlier (not used PV3) \\
ID\_MEAS\_SKIP\_PHOTOM       & 0x00000004 & detection was ignored for photometry measurement (not used PV3) \\
ID\_MEAS\_AREA               & 0x00000008 & detection near image edge (not used PV3) \\
ID\_MEAS\_POOR\_ASTROM       & 0x00000010 & detection is astrometry outlier \\
ID\_MEAS\_SKIP\_ASTROM       & 0x00000020 & detection was ignored for astrometry measurement \\
ID\_MEAS\_USED\_OBJ          & 0x00000040 & detection was used during update objects \\
ID\_MEAS\_USED\_CHIP         & 0x00000080 & detection was used during update chips (not saved PV3) \\
ID\_MEAS\_BLEND\_MEAS        & 0x00000100 & detection is within radius of multiple objects \\
ID\_MEAS\_BLEND\_OBJ         & 0x00000200 & multiple detections within radius of object \\
ID\_MEAS\_WARP\_USED         & 0x00000400 & measurement used to find mean warp photometry \\
ID\_MEAS\_UNMASKED\_ASTRO    & 0x00000800 & measurement was not masked in final astrometry fit \\
ID\_MEAS\_BLEND\_MEAS\_X     & 0x00001000 & detection is within radius of multiple objects across catalogs \\
ID\_MEAS\_ARTIFACT           & 0x00002000 & detection is thought to be non-astronomical \\
ID\_MEAS\_SYNTH\_MAG         & 0x00004000 & magnitude is synthetic \\
ID\_MEAS\_PHOTOM\_UBERCAL    & 0x00008000 & externally-supplied zero point from ubercal analysis \\
ID\_MEAS\_STACK\_PRIMARY     & 0x00010000 & this stack measurement is in the primary skycell \\
ID\_MEAS\_STACK\_PHOT\_SRC   & 0x00020000 & this measurement supplied the stack photometry \\
ID\_MEAS\_ICRF\_QSO          & 0x00040000 & this measurement is an ICRF reference position \\
ID\_MEAS\_IMAGE\_EPOCH       & 0x00080000 & this measurement is registered to the image epoch (not tied to ref catalog epoch) \\
ID\_MEAS\_PHOTOM\_PSF        & 0x00100000 & this measurement is used for the mean psf mag \\
ID\_MEAS\_PHOTOM\_APER       & 0x00200000 & this measurement is used for the mean ap mag \\
ID\_MEAS\_PHOTOM\_KRON       & 0x00400000 & this measurement is used for the mean kron mag \\
ID\_MEAS\_MASKED\_PSF        & 0x01000000 & this measurement is masked based on IRLS weights for mean psf mag \\
ID\_MEAS\_MASKED\_APER       & 0x02000000 & this measurement is masked based on IRLS weights for mean ap mag \\
ID\_MEAS\_MASKED\_KRON       & 0x04000000 & this measurement is masked based on IRLS weights for mean kron mag \\
ID\_MEAS\_OBJECT\_HAS\_2MASS & 0x10000000 & measurement comes from an object with 2mass data \\
ID\_MEAS\_OBJECT\_HAS\_GAIA  & 0x20000000 & measurement comes from an object with gaia data \\
ID\_MEAS\_OBJECT\_HAS\_TYCHO & 0x40000000 & measurement comes from an object with tycho data \\
\hline
\end{tabular}
\end{center}
\end{table*}

\begin{table*}
\begin{center}
\footnotesize
\caption{\label{tab:secf_mask_values} Relphot Per-Filter Info Flag Bit Values} 
\begin{tabular}{lcl}
\hline
\hline
{\bf Bit Name} & {\bf Bit Value} & {\bf Description} \\
\hline
ID\_SECF\_STAR\_FEW    		   & 0x00000001 & Used within relphot: skip star \\
ID\_SECF\_STAR\_POOR   		   & 0x00000002 & Used within relphot: skip star \\
ID\_SECF\_USE\_SYNTH   		   & 0x00000004 & Synthetic photometry used in average measurement \\
ID\_SECF\_USE\_UBERCAL 		   & 0x00000008 & Ubercal photometry used in average measurement \\
ID\_SECF\_HAS\_PS1     		   & 0x00000010 & PS1 photometry used in average measurement \\
ID\_SECF\_HAS\_PS1\_STACK 	   & 0x00000020 & PS1 stack photometry exists \\
ID\_SECF\_HAS\_TYCHO   		   & 0x00000040 & Tycho photometry used for synth mags \\
ID\_SECF\_FIX\_SYNTH   		   & 0x00000080 & Synth mags repaired with zpt map \\
ID\_SECF\_RANK\_0    		   & 0x00000100 & Average magnitude uses rank 0 values \\
ID\_SECF\_RANK\_1    		   & 0x00000200 & Average magnitude uses rank 1 values \\
ID\_SECF\_RANK\_2    		   & 0x00000400 & Average magnitude uses rank 2 values \\
ID\_SECF\_RANK\_3    		   & 0x00000800 & Average magnitude uses rank 3 values \\
ID\_SECF\_RANK\_4    		   & 0x00001000 & Average magnitude uses rank 4 values \\
ID\_SECF\_OBJ\_EXT\_PSPS  	   & 0x00002000 & In PSPS ID\_SECF\_OBJ\_EXT is saved here so it fits within 16 bits  \\
ID\_SECF\_STACK\_PRIMARY 	   & 0x00004000 & PS1 stack photometry includes a primary skycell \\
ID\_SECF\_STACK\_BESTDET 	   & 0x00008000 & PS1 stack best measurement is a detection (not forced) \\
ID\_SECF\_STACK\_PRIMDET 	   & 0x00010000 & PS1 stack primary measurement is a detection (not forced) \\
ID\_SECF\_STACK\_PRIMARY\_MULTIPLE & 0x00020000 & PS1 stack object has multiple primary measurements \\
ID\_SECF\_HAS\_SDSS      	   & 0x00100000 & This photcode has SDSS photometry \\
ID\_SECF\_HAS\_HSC       	   & 0x00200000 & This photcode has HSC  photometry \\
ID\_SECF\_HAS\_CFH       	   & 0x00400000 & This photcode has CFH  photometry (mostly Megacam) \\
ID\_SECF\_HAS\_DES       	   & 0x00800000 & This photcode has DES  photometry \\
ID\_SECF\_OBJ\_EXT       	   & 0x01000000 & Extended in this band \\
\hline
\end{tabular}
\end{center}
\end{table*}

\begin{table*}
\begin{center}
\footnotesize
\caption{\label{tab:object_mask_values} Per-Object Flag Bit Values} 
\begin{tabular}{lcl}
\hline
\hline
{\bf Bit Name} & {\bf Bit Value} & {\bf Description} \\
\hline
ID\_OBJ\_FEW               & 0x00000001 & used within relphot: skip star \\
ID\_OBJ\_POOR              & 0x00000002 & used within relphot: skip star \\
ID\_OBJ\_ICRF\_QSO         & 0x00000004 & object IDed with known ICRF quasar (may have ICRF position measurement) \\
ID\_OBJ\_HERN\_QSO\_P60    & 0x00000008 & identified as likely QSO \citep{2016ApJ...817...73H}, $P_{\rm QSO} \geq 0.60$ \\
ID\_OBJ\_HERN\_QSO\_P05    & 0x00000010 & identified as possible QSO \citep{2016ApJ...817...73H}, $P_{\rm QSO} \geq 0.05$ \\
ID\_OBJ\_HERN\_RRL\_P60    & 0x00000020 & identified as likely  RR Lyra \citep{2016ApJ...817...73H}, $P_{\rm RRLyra} \geq 0.60$ \\
ID\_OBJ\_HERN\_RRL\_P05    & 0x00000040 & identified as possible RR Lyra \citep{2016ApJ...817...73H}, $P_{\rm RRLyra} \geq 0.05$ \\
ID\_OBJ\_HERN\_VARIABLE    & 0x00000080 & identified as a variable by \cite{2016ApJ...817...73H} \\
ID\_OBJ\_TRANSIENT         & 0x00000100 & identified as a non-periodic (stationary) transient \\
ID\_OBJ\_HAS\_SOLSYS\_DET  & 0x00000200 & identified with a known solar-system object (asteroid or other) \\
ID\_OBJ\_MOST\_SOLSYS\_DET & 0x00000400 & most detections from a known solar-system object \\
ID\_OBJ\_LARGE\_PM         & 0x00000800 & star with large proper motion \\
ID\_OBJ\_RAW\_AVE      	   & 0x00001000 & simple weighted average position was used (no IRLS fitting) \\
ID\_OBJ\_FIT\_AVE      	   & 0x00002000 & average position was fitted \\
ID\_OBJ\_FIT\_PM       	   & 0x00004000 & proper-motion model was fitted \\
ID\_OBJ\_FIT\_PAR      	   & 0x00008000 & full parallax and proper-motion model was fitted \\
ID\_OBJ\_USE\_AVE      	   & 0x00010000 & average position used (no proper-motion or parallax) \\
ID\_OBJ\_USE\_PM       	   & 0x00020000 & proper motion fit used (no parallax) \\
ID\_OBJ\_USE\_PAR      	   & 0x00040000 & full fit with proper motion and parallax \\
ID\_OBJ\_NO\_MEAN\_ASTROM  & 0x00080000 & mean astrometry could not be measured \\
ID\_OBJ\_STACK\_FOR\_MEAN  & 0x00100000 & stack position used for mean astrometry \\
ID\_OBJ\_MEAN\_FOR\_STACK  & 0x00200000 & mean astrometry could not be measured \\
ID\_OBJ\_BAD\_PM           & 0x00400000 & failure to measure proper-motion model \\
ID\_OBJ\_EXT               & 0x00800000 & extended in Pan-STARRS data \\
ID\_OBJ\_EXT\_ALT          & 0x01000000 & extended in external data (2MASS) \\
ID\_OBJ\_GOOD              & 0x02000000 & good-quality measurement in Pan-STARRS data \\
ID\_OBJ\_GOOD\_ALT         & 0x04000000 & good-quality measurement in  external data (2MASS) \\
ID\_OBJ\_GOOD\_STACK       & 0x08000000 & good-quality object in the stack ($> 1$ good stack) \\
ID\_OBJ\_BEST\_STACK       & 0x10000000 & the primary stack measurements are the ``best'' measurements \\
ID\_OBJ\_SUSPECT\_STACK    & 0x20000000 & suspect object in the stack ($> 1$ good or suspect stack, $< 2$ good) \\
ID\_OBJ\_BAD\_STACK        & 0x40000000 & poor-quality object in the stack ($< 1$ good stack) \\
\hline
\end{tabular}
\end{center}
\end{table*}


\begin{table*}
\begin{center}
\footnotesize
\caption{\label{tab:image_mask_values} Per-Image Flag Bit Values} 
\begin{tabular}{lcl}
\hline
\hline
{\bf Bit Name} & {\bf Bit Value} & {\bf Description} \\
\hline
ID\_IMAGE\_NEW             & 0x00000000 & no calibrations yet attempted \\
ID\_IMAGE\_PHOTOM\_NOCAL   & 0x00000001 & user-set value used within relphot: ignore \\
ID\_IMAGE\_PHOTOM\_POOR    & 0x00000002 & relphot says image is bad (dMcal > limit) \\
ID\_IMAGE\_PHOTOM\_SKIP    & 0x00000004 & user-set value: assert that this image has bad photometry \\
ID\_IMAGE\_PHOTOM\_FEW     & 0x00000008 & currently too few measurements for photometry \\
ID\_IMAGE\_ASTROM\_NOCAL   & 0x00000010 & user-set value used within relastro: ignore \\
ID\_IMAGE\_ASTROM\_POOR    & 0x00000020 & relastro says image is bad (dR,dD > limit) \\
ID\_IMAGE\_ASTROM\_FAIL    & 0x00000040 & relastro fit diverged, fit not applied \\
ID\_IMAGE\_ASTROM\_SKIP    & 0x00000080 & user-set value: assert that this image has bad astrometry \\
ID\_IMAGE\_ASTROM\_FEW     & 0x00000100 & currently too few measurements for astrometry \\
ID\_IMAGE\_PHOTOM\_UBERCAL & 0x00000200 & externally-supplied photometry zero point from ubercal analysis \\
ID\_IMAGE\_ASTROM\_GMM     & 0x00000400 & image was fitted to positions corrected by the galaxy motion model \\
\hline
\end{tabular}
\end{center}
\end{table*}


\section{Photometry Calibration}

\subsection{Ubercal Analysis}


The photometric calibration of the DVO database starts with the
``ubercal'' analysis technique as described by
\cite{2012ApJ...756..158S}.  This analysis is performed by the group
at Harvard, loading data from the \code{smf} files into their instance
of the Large Scale Database \citep[LSD,][]{2011AAS...21743319J}, a
system similar to DVO used to manage the detections and determine the
calibrations.

Photometric nights are selected and all other exposures are ignored.
Each night is allowed to have a single fitted zero point
(corresponding to the sum $zp_{\rm nominal} + M_{cal}$ below) and a
single fitted value for the airmass extinction coefficient ($K_{\rm
  \lambda}$) per filter.  The zero points and extinction terms are
determined as a least squares minimization process using the repeated
measurements of the same stars from different nights to tie nights
together.  Flat-field corrections are also determined as part of the
minimization process.  In the original (PV1) ubercal analysis,
\cite{2012ApJ...756..158S} determined flat-field corrections for
$2\times 2$ sub-regions of each chip in the camera and four distinct
time periods (``seasons'').  Later analysis (PV2) used an $8\times8$
grid of flat-field corrections to good effect.

The ubercal analysis was re-run for PV3 by the Harvard group.  For the
PV3 analysis, under the pressure of time to complete the analysis, we
chose to use only a $2\times 2$ grid per chip as part of the ubercal
fit and to leave higher frequency structures to the later analysis.  A
5th flat-field season consisting of nearly the last 2 years of data
was also included for PV3.  In retrospect, as we show below, the data
from the latter part of the survey would probably benefit from
additional flat-field seasons. 


By excluding non-photometric data and only fitting 2 parameters for
each night, the Ubercal solution is robust and rigid.  It is not
subject to unexpected drift or sensitivity of the solution to the
vagaries of the data set.  The Ubercal analysis is also especially
aided by the inclusion of multiple Medium Deep field observations
every night, helping to tie down overall variations of the system
throughput and acting as internal standard star fields.  The resulting
photometric system is shown by \cite{2012ApJ...756..158S} to have reliability
across the survey region at the level of (8.0, 7.0, 9.0, 10.7, 12.4)
millimags in (\grizy).  As we discuss below, this conclusion is
reinforced by our external comparison.  


The overall zero point for each filter is not naturally determined by
the Ubercal analysis; an external constraint on the overall
photometric system is required for each filter.
\cite{2012ApJ...756..158S} used photometry of the MD09 Medium Deep
field to match the photometry measured by \cite{2012ApJ...750...99T}
on the reference photometric night of MJD 55744 (UT 02 July 2011).
\cite{2014ApJ...795...45S} and \cite{2015ApJ...815..117S} have
re-examined the photometry of Calspec standards \citep{1996AJ....111.1743B} as
observed by PS1.  \cite{2014ApJ...795...45S} reject 2 of the 7 stars
used by \cite{2012ApJ...750...99T} and add photometry of 5 additional
stars.  \cite{2015ApJ...815..117S} further reject measurements of
Calspec standards obtained close to the center of the camera field of
view where the PSF size and shape changes very rapidly.  The result of
this analysis modifies the over system zero points by 20 - 35
millimags compared with the system determined by
\cite{2012ApJ...756..158S}.



\subsection{Applying the Ubercal Zero Points : Setphot}

The ubercal analysis above results in a table of zero points for all
exposures considered to be photometric, along with a set of
low-resolution flat-field corrections.  It is now necessary to use this
information to determine zero points for the remaining exposures and
to improve the resolution of the flat-field correction.  This analysis
is done within the IPP DVO database system.

The ubercal zero points and the flat-field correction data are loaded
into the PV3 DVO database using the program \code{setphot}.  This
program converts the reported zero point and flat field values to the
DVO internal representation in which the zero point of each image is
split into three main components:
\begin{equation} 
zp_{\rm total} = zp_{\rm nominal} + M_{cal} + K_{\rm \lambda}(\sec \zeta - 1)
\end{equation}
where $zp_{\rm nominal}$ and $K_{\rm \lambda}$ are static values for
each filter representing respectively the nominal zero point and the
slope of the trend with respect to the airmass ($\zeta$) for each
filter.  These static values are listed in Table~\ref{tab:zpts}.  When
\code{setphot} was run, these static zero points have been adjusted by
the Calspec offsets listed in Table~\ref{tab:zpts} based on the
analysis of Calspec standards by \cite{2015ApJ...815..117S}.  These
offsets bring the photometric system defined by the ubercal analysis
into alignment with \cite{2015ApJ...815..117S}.  The value $M_{cal}$
is the offset needed by each exposure to match the ubercal value, or
to bring the non-ubercal exposures into agreement with the rest of the
exposures, as discussed below.  The flat-field information is encoded
in a table of flat-field offsets as a function of time, filter, and
camera position.  Each image which is part of the ubercal subset is
marked with a bit in the field \code{Image.flags}:
\code{ID_IMAGE_PHOTOM_UBERCAL = 0x00000200}.  

\begin{table}[hb]
\begin{center}
\caption{PS1 / GPC1 Zero Points and Coefficients\label{tab:zpts}}
\begin{tabular}{llll}
\hline
\hline
{\bf Filter} & {\bf Zero Point} & {\bf Zero Point} & {\bf Airmass Slope} \\
 & {\bf (Raw)} & {\bf (Calspec)} & \\
\hline
\gps & 24.563 & 24.583 & 0.147 \\
\rps & 24.750 & 24.783 & 0.085 \\
\ips & 24.611 & 24.635 & 0.044 \\
\zps & 24.240 & 24.278 & 0.033 \\
\yps & 23.320 & 23.331 & 0.073 \\
\hline
\end{tabular}
\end{center}
\end{table}


When \code{setphot} applies the ubercal information to the image
tables, it also updates the individual measurements associated with
those images.  In the DVO database schema, the normalized instrumental
magnitude, $M_{\rm inst} = -2.5log_{10} (DN / sec)$ is stored 
for each measurement, with an arbitrary (but fixed)
constant offset of 25 to place the modified instrumental magnitudes into
approximately the correct range.  Associated with each measurement are
two correction magnitudes: $M_{\rm cal}$ and $M_{\rm flat}$, along
with the airmass for the measurement, calculated using the altitude of
the individual detection as determined from the Right Ascension,
Declination, the observatory latitude, and the sidereal time.  For a
camera with the field of view of the PS1 GPC1, the airmass may vary
significantly within the field of view, especially at low elevations.
In the worst cases, at the celestial pole, the airmass within a single
exposure may span a range of 2.56 - 2.93.  The complete calibrated
(`relative') magnitude is determined from the stored database values
as:
\begin{equation}
M_{\rm rel} = M_{\rm inst} + zp_{\rm ref} + M_{\rm cal} + M_{\rm flat} + K_\lambda (sec \zeta - 1).
\end{equation}
The calibration offsets, $M_{\rm cal}$ and $M_{\rm flat}$, represent
the per-exposure zero point correction and the slowly-changing
flat-field correction respectively.  These two values are split so the
flat-field corrections may be determined and applied independently
from the time-resolved zero point variations.  Note that the above
corrections are applied to each of the types of measurements stored in
the database, PSF, Aperture, Kron.  The calibration math remains the
same regardless of the kind of magnitude being measured.  Also note
that for the moment, this discussion should only be considered as
relevant to the chip measurements.  Below we discuss the implications
for the stack and warp measurements.

When the ubercal zero points and flat-field data are loaded,
\code{setphot} updates the $M_{\rm cal}$ values for all measurements
which have been derived from the ubercal images.  These measurements
are also marked in the field \code{Measure.dbFlags} with the bit
\code{ID_MEAS_PHOTOM_UBERCAL = 0x00008000}.  At this stage,
\code{setphot} also updates the values of $M_{\rm flat}$ for all GPC1
measurements in the appropriate filters.

\subsection{Relphot Analysis}


Relative photometry is used to determine the zero points of the
exposures which were not included in the ubercal analysis.  The
relative photometry analysis has been described in the past by
\cite{2013ApJS..205...20M}.  We review that analysis here, along with
specific updates for PV3.

As described above, the instrumental magnitude and the calibrated magnitude
are related by arithmetic magnitude offsets which account for effects
such as the instrumental variations and atmospheric attenuation.  
\begin{equation}
M_{rel} = m_{inst} + ZP + M_{cal}
\end{equation}

From the collection of measurements, we can generate an average
magnitude for a single star (or other object):
\begin{equation}
  M_{ave} = \frac{\sum_i M_{rel,i} w_i}{\sum_i w_i}
\end{equation}
We find that the color difference of the different chips can be
ignored, and set the color-trend slope to 0.0.  Note that we only use
a single mean airmass extinction term for all exposures -- the
difference between the mean and the specific value for a given night
is taken up as an additional element of the atmospheric attenuation.


We write a global $\chi^2$ equation which we attempt to minimize by
finding the best mean magnitudes for all objects and the best
$M_{\rm cal}$ offset for each exposure:
\begin{equation}
  \chi^2 = \frac{\sum_{i,j} (m_{inst}[i,j] + ZP + K \zeta +
    M_{clouds}[i] - M_{ave}[j]) w_{i,j}}{\sum_{i,j} w_{i,j}}
\end{equation}

If everything were fitted at once and allowed to float, this system of
equations would have $N_{exposures} + N_{stars} \sim 2 \times 10^5 + N
\times 10^9$ unknowns.  We solve the system of equations by iteration,
solving first for the best set of mean magnitudes in the assumption of
zero clouds, then solving for the clouds implied by the differences
from these mean magnitudes.  Even with 1-2 magnitudes of extinction,
the offsets converge to the milli-magnitude level within 8 iterations.

Only brighter, high quality measurements are used in the relative
photometry analysis of the exposure zero points.  We use only the
brighter objects, limiting the density to a maximum of 4000 objects
per square degree (lower in areas where we have more observations).
When limiting the density, we prefer objects which are brighter (but
not saturated), and those with the most measurements (to ensure better
coverage over the available images).

There are a few classes of outliers which we need to be careful to
detect and avoid.  First, any single measurement may be deviant for a
number of reasons (e.g., it lands in a bad region of the detector,
contamination by a diffraction spike or other optical artifact, etc).
We attempt to exclude these poor measurements in advance by rejecting
measurements which the photometric analysis has flagged the result as
suspicious.  We reject detections which are excessively masked; these include
detections which are too close to other bright objects, diffraction
spikes, ghost images, or the detector edges.  However, these
rejections do not catch all cases of bad measurements.


After the initial iterations, we also perform outlier rejections based
on the consistency of the measurements.  For each star, we use a two
pass outlier clipping process.  We first define a robust median and
sigma from the inner 50\% of the measurements.  Measurements which are
more than 5$\sigma$ from this median value are rejected, and the mean
\& standard deviation (weighted by the inverse error) are
recalculated.  We then reject detections which are more than 3$\sigma$
from the recalculated mean.  

Suspicious stars are also excluded from the analysis.  We exclude stars
with reduced $\chi^2$ values more than 20.0, or more than 2$\times$
the median, whichever is larger.  We also exclude stars with standard
deviation (of the measurements used for the mean) greater than 0.005
mags or 2$\times$ the median standard deviation, whichever is greater.


Similarly for images, we exclude those with more than 2 magnitudes of
extinction or for which the deviation greater of the zero points per
star are than 0.075 mags or 2$\times$ the median value, whichever is
greater.  These cuts are somewhat conservative to limit us to only
good measurements.  The images and stars rejected above are not used
to calculate the system of zero points and mean magnitudes.  These
cuts are updated several times as the iterations proceed.  After the
iterations have completed, the images which have been reject are
calibrated based on their overlaps with other images.

We overweight the ubercal measurements in order to tie the relative
photometry system to the ubercal zero points.  Ubercal images and
measurements from those images are not allowed to float in the
relative photometry analysis.  Detections from the Ubercal images are
assigned weights of 10x their default (inverse-variance) weight.  The
calculation of the formal error on the mean magnitudes propagates this
additional weight, so that the errors on the Ubercal observations
dominates where they are present. 

\begin{equation}
  \mu = \frac{\sum m_i w_i \sigma_i^{-2}}{\sum w_i \sigma_i^{-2}}
\end{equation}
\begin{equation}
  \sigma_\mu = \frac{\sum w_i^2 \sigma_i^{-2}}{(\sum w_i
    \sigma_i^{-2})^2}
\end{equation}

The calculation of the relative photometry zero points is performed
for the entire $3\pi$ data set in a single, highly parallelized
analysis.  As discussed above, the measurement and object data in the
DVO database are distributed across a large number of computers in the
IPP cluster: for PV3, 100 parallel hosts are used.  These machines by
design control data from a large number of unconnected small patches
on the sky, with the goal of speeding queries for arbitrary regions of
the sky.  As a result, this parallelization is entirely inappropriate
as the basis of the relative photometry analysis.  For the relative
photometry calculation (and later for relative astrometry
calculation), the sky is divided into a number of large, contiguous
regions each bounded by lines of constant RA \& DEC, 73 regions in the
case of the PV3 analysis.  A separate computer, called a ``region
host'' is responsible for each of these regions: that computer is
responsible for calculating the mean magnitudes of the objects which
land within its region and for determining the exposure zero points
for exposures for which the center of the exposure lands in the region
of responsibility.  

\begin{figure*}[htbp]
 \begin{center}
  \begin{minipage}{0.85\linewidth}
   \includegraphics[width=\textwidth,clip]{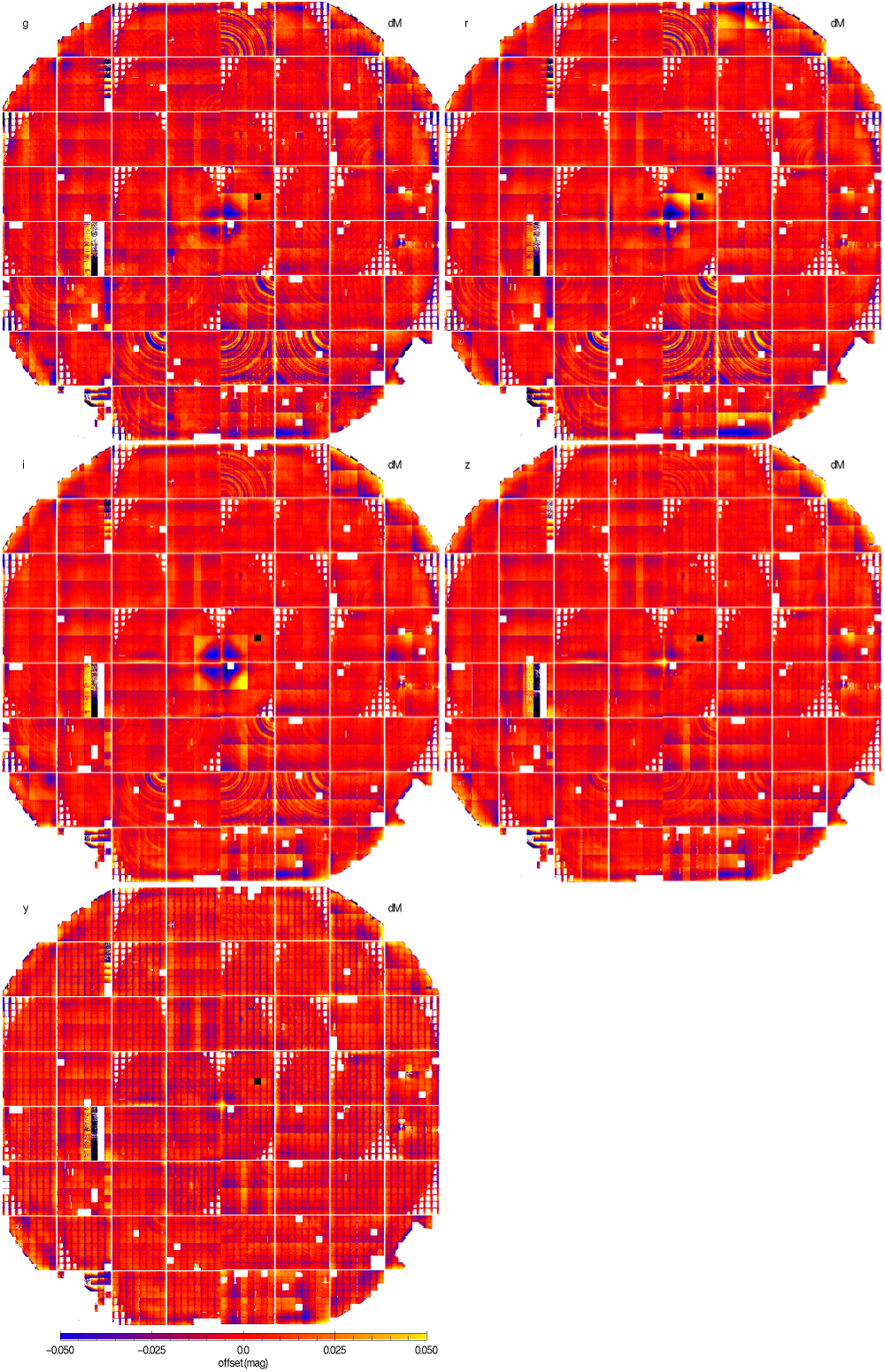}
  \end{minipage}
  \hspace{-2.75in}
  \begin{minipage}{0.4\linewidth}
   \vspace{3.25in}
   \caption{\label{fig:photflat} High-resolution flat-field correction images for the 5 filters $grizy$.}
  \end{minipage}
 \end{center}
\end{figure*}

The iterations described above (calculate mean
magnitudes, calculate zero points, calculate new measurements) are
performed on each of the 73 region hosts in parallel.  However, between
certain iteration steps, the region hosts must share some information.
After mean object magnitudes are calculated, the region hosts must
share the object magnitudes for the objects which are observed by
exposures controlled by neighboring region hosts.  After image
calibrations have been determined by each region host, the image
calibrations must be shared with the neighboring region hosts so
measurement values associated with objects owned by a neighboring
region host may be updated.

The complete work flow of the all-sky relative photometry analysis
starts with an instance of the program running on a master computer.
This machine loads the image database table and assigns the images to
the 73 region hosts.  A process is then launched on each of the region
hosts which is responsible for managing the image calibration analysis
on that host.  These processes in turn make an initial request of the
photometry information (object and measurement) from the 100 parallel
DVO partition machines.  In practice, the processes on the the region
hosts are launched in series by the master process to avoid
overloading the DVO partition machines with requests for photometry
data from all region hosts at once.  Once all of the photometry has
been loaded, the region hosts perform their iterations, sharing the
data which they need to share with their neighbors and blocking while
they wait for the data they need to receive from their neighbors.  The
management of this stage is performed by communication between the
region host.  At the end of the iterations, the regions hosts write out
their final image calibrations.  The master machine then loads the
full set of image calibrations and then applies these calibrations
back to all measurements in the database, updating the mean photometry
as part of this process.  The calculations for this last step are
performed in parallel on the DVO partition machines.

With the above software, we are able to perform the entire relphot
analysis for the full 3$\pi$ region at once, avoiding any possible
edge effects.  The region host machines have internal memory ranging
from 96GB to 192GB.  Regions are drawn, and the maximum allowed
density was chosen, to match the memory usage to the memory available
on each machine.  A total of 9.8TB of RAM was available for the
analysis, allowing for up to 6000 objects per square degree in the
analysis.

\begin{figure*}[htbp]
  \begin{center}
 \includegraphics[height=\vsize,clip]{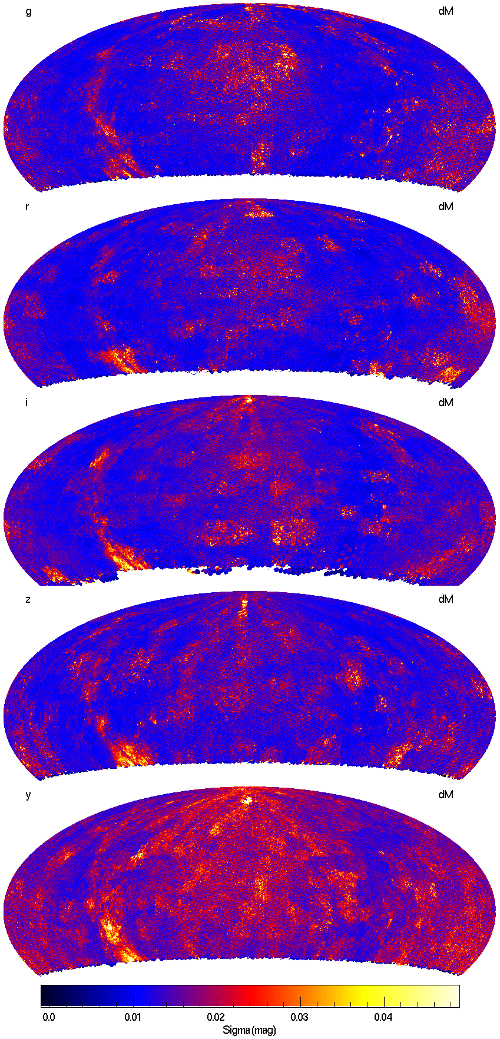}
  \caption{\label{fig:allsky.photom.sigma} Consistency of photometry
    measurements across the sky.  Each panel shows a map of the
    standard deviation of photometry residuals for stars in each
    pixel.  The median value of the measure standard deviations across
    the sky is $(\sigma_g, \sigma_r, \sigma_i, \sigma_z, \sigma_y) =
    (14, 14, 15, 15, 18)$ millimags.  These values reflect the typical
    single-measurement errors for bright stars.}
  \end{center}
\end{figure*}


\subsubsection{Photometric Flat-field}
\label{sec:phot.flat}

For PV3, the relphot analysis was performed two times.  The first
analysis used only the flat-field corrections determined by the
ubercal analysis, with a resolution of 2x2 flat-field values for each
GPC1 chip (corresponding to \approx 2400 pixels), and 5 separate
flat-field 'seasons'.  However, we knew from prior studies that there
were significant flat-field structures on smaller scales.  We used the
data in DVO after the initial relphot calibration to measure the
flat-field residual with much finer resolution: 124 x 124 flat-field
values for each GPC1 chip (40x40 pixels per point).  We then used
\code{setphot} to apply this new flat-field correction, as well as the
ubercal flat-field corrections, to the data in the database.  At this
point, we re-ran the entire relphot analysis to determine zero points
and to set the average magnitudes.

Figure~\ref{fig:photflat} shows the high-resolution photometric
flat-field corrections applied to the measurements in the DVO
database.  These flat-fields make low-level corrections of up to
\approx 0.03 magnitudes.  Several features of interest are apparent in
these images.  

First, at the center of the camera is an important structure caused by
the telescope optics which we call the ``tent''.  In this portion of
the focal plane, the image quality degrades very quickly.  The
photometry is systematically biased because the point spread function
model cannot follow the real changes in the PSF shape on these small
scales.  As is evident in the image, the effect is such that the flux
measured using a PSF model is systematically low, as expected if the
PSF model is too small.  

The square outline surrounding the ``tent'' is due to the 2$\times$2
sampling per chip used for the Ubercal flat-field corrections.  The
imprint of the Ubercal flat-field is visible throughout this
high-resolution flat-field: in regions where the underlying flat-field
structure follows a smooth gradient across a chip, the Ubercal
flat-field partly corrects the structure, leaving behind a saw-tooth
residual.  The high-resolution flat-field corrects the residual
structures well.

Especially notable in the bluer filters is a pattern of quarter
circles centered on the corners of the chips.  These patterns are
similar to the ``tree rings'' reported by the Dark Energy Survey team
\citep{2014PASP..126..750P} and identified as a result of lateral
migration of electrons in the detectors due to electric fields due to
dopant variations.  Unlike the tree ring features discussed by these
other authors, the strong features observed in the GPC1 photometry are
not caused by lateral electric fields, but rather by variations in the
vertical electron diffusion rate due to electric field variations
perpendicular to the plane of the detector.  This effect is discussed
in detail by \cite{2018PASP..130f5002M}.  The photometric features are
due to low-level changes in the PSF size which we attribute to the
variable charge diffusion.

Other features include some poorly responding cells (e.g., in XY14)
and effects at the edges of chips, possibly where the PSF model fails
to follow the changes in the PSF.



For stacks and warps, the image calibrations were determined after the
relative photometry was performed on the individual chips.  Each stack
and each warp was tied via relative photometry to the average
magnitudes from the chip photometry.  In this case, no flat-field
corrections were applied.  For the stacks, such a correction would not
be possible after the stack has been generated because multiple chip
coordinates contribute to each stack pixel coordinate.  For the warps,
it is in principle possible to map back to the corresponding chip, but
the information was not available in the DVO database, and thus it was
not possible at this time to determine the flat-field correction
appropriate for a given warp.  This latter effect is one of several
which degrade the warp photometry compared to the chip photometry at
the bright end.

For the stack calibration, we calculate two separate zero points: one
for photometry tied to the PSF model and a second for the
aperture-like measurements (total aperture magnitudes, Kron magnitude,
circular fixed-radius aperture magnitudes).  This split is needed
because of the limited quality of the stack PSF photometry due to the
highly variable PSF in the stacks.  Aperture magnitudes, however, are
not significantly affected by the PSF variations.  We therefore tie
the PSF magnitudes to the average of the chip photometry PSF
magnitudes, but the aperture-like magnitudes are tied by equating the
stack Kron magnitudes to the average chip Kron magnitudes.  {\em Note
  that for DR1, this split zero point calibration was used; instead
  all stack photometry was tied to the average chip photometry via the
  PSF magnitudes.}  The result of using a single zero point is that
the stack PSF magnitudes are consistent across the sky with the chip
PSF magnitudes, but the aperture-like magnitudes show significant
spatial variations.  Figure~\ref{fig:stack.bad.kron} illustrates the
impact of using a single PSF zero point for the stack photometry.
This split is not needed for the forced-warp photometry since the
individual warps have well-defined PSfs.

\subsection{Photometry Calibration Quality}

Figure~\ref{fig:allsky.photom.sigma} shows the standard deviations of
the mean residual photometry for bright stars as a function of
position across the sky.  For each pixel in these images, we selected
all objects with (14.5, 14.5, 14.5, 14.0, 13.0) $<$ ($g,r,i,z,y$) $<$
(17, 17, 17, 16.5, 15.5), with at least 3 measurements in $i$-band (to
reject artifacts detected in a pair of exposures from the same night),
with \code{PSF_QF} $> 0.85$ (to reject excessively-masked objects),
and with $mag_{\rm PSF} - mag_{\rm Kron} < 0.1$ (to reject galaxies).
We then generated histograms of the difference between the average
magnitude and the apparent magnitude in an individual image for each
filter for all stars in a given pixel in the images.  From these
residual histograms, we can then determine the median and the 68\%-ile
range to calculate a robust standard deviation.  This represents the
bright-end systematic error floor for a measurement from a single
exposure.  The standard deviations are then plotted in
Figure~\ref{fig:allsky.photom.sigma}.  

The 5 panels in Figure~\ref{fig:allsky.photom.sigma} show several
features.  The Galactic bulge is clearly seen in all five filters,
with the impact strongest in the reddest bands.  We attribute this to
the effects of crowding and contamination of the photometry by
neighbors.  Large-scale, roughly square features \approx 10 degrees on
a side in these images can be attributed to the vagaries of weather:
these patches correspond to the observing chunks.  These images
include both photometric and non-photometric exposures.  It seems
plausible that the non-photometric images from relatively poor quality
nights elevate the typical errors.  On small scales, there are
circular patterns \approx 3 degrees in diameter corresponding to
individual exposures; these represent residual flat-fields structures
not corrected by our stellar flat-fielding.  The median of the
standard deviations in the five filters are
$(\sigma_g,\sigma_r,\sigma_i,\sigma_z,\sigma_y) = (14, 14, 15, 15,
18)$ millimagnitudes.


\subsection{Calculation of Object Photometry}

Once the image photometric calibrations (zero points and flat-field
corrections) have been determined and applied to the measurements from
each image, we can calculate the best average photometry for each
object.  We calculate average magnitudes for the chip photometry; for
the forced-warp photometry, we calculate the average of the fluxes and
report both average fluxes and the equivalent average magnitudes.
Since the chip photometry requires signal-to-noise of 5 for a
detection, the bias introduced by averaging magnitudes is small.
Since the forced-warp photometry measurements are low signal-to-noise,
with potentially negative flux values, it is necessary to average the
fluxes.

The first challenge is to select which measurements to use in
the calculation of the average photometry.  For the $3\pi$ Survey
data, a single object may have anywhere from zero to roughly twenty
measurements in a given filter.  Not all measurements are of equal
value, but we need a process which assigns an average photometry value
in all cases (and a way for the user to recognize average values which
should be treated with care).  As discussed in more detail below, we
have defined a triage process to select the ``best'' set of
measurements available in each filter for each object.  Once the set
of measurements to be used in the analysis is determined, we use the
Iteratively Reweighted Least Squares (IRLS) technique to determine the
average photometry given the possible presence of non-Gaussian
outliers even within the best subset of measurements.  

\subsubsection{Selection of Measurements}

To choose the measurements which will be used in the analysis, we 
give each measurement a rank value based on a variety of tests of the
quality of the measurement, with lower values being better quality.
In the description below
The ranking values are defined as follows:
\begin{itemize}
\item {\bf rank 0 :} perfect measurement (no quality concerns)
\item {\bf rank 1 :} PSF ``perfect pixel'' quality factor (\code{PSF_QF_PERFECT}) $< 0.85$.  \code{PSF_QF_PERFECT} measures the PSF-weighted fraction of pixels which are not masked \citep[see][]{magnier2017.analysis}.
\item {\bf rank 2 :} Photometry analysis flag field (\code{photFlags}) has one of the ``poor quality'' bits raised.  These bits are listed below; OR-ed together they have the hexadecimal value \code{0xe0440130}
\begin{itemize}
  \item {\tt PM\_SOURCE\_MODE\_POOR = 0x00000010} : Fit succeeded, but with low-S/N or high-Chisq 
  \item {\tt PM\_SOURCE\_MODE\_PAIR = 0x00000020} : Source fitted with a double psf
  \item {\tt PM\_SOURCE\_MODE\_BLEND = 0x00000100} : Source is a blend with other sources
  \item {\tt PM\_SOURCE\_MODE\_BELOW\_MOMENTS\_SN = 0x00040000} : Moments not measured due to low S/N
  \item {\tt PM\_SOURCE\_MODE\_BLEND\_FIT = 0x00400000} : Source was fitted as a blended object
  \item {\tt PM\_SOURCE\_MODE\_ON\_SPIKE = 0x20000000} : Peak lands on diffraction spike
  \item {\tt PM\_SOURCE\_MODE\_ON\_GHOST = 0x40000000} : Peak lands on ghost or glint
  \item {\tt PM\_SOURCE\_MODE\_OFF\_CHIP = 0x80000000} : peak lands off edge of chip
\end{itemize}
\item {\bf rank 3 :} Poor measurement as defined by relphot.  This may be due to a fixed allowed region on the detector, or due to an outlier clipped analysis.  In the $3\pi$ PV3 calibration, these tests were not applied.
\item {\bf rank 4 :} PSF quality factor (\code{PSF_QF}) $< 0.85$.
  \code{PSF_QF} measures the PSF-weighted fraction of pixels which are
  not masked as ``bad'', but may be ``suspect''.  Bad values are
  blank, highly non-linear or non-responsive; suspect pixels include
  those pixels on ghosts, diffraction spikes, bright star bleeds, and
  the mildly-saturated cores of bright stars.  Suspect values may have
  some use in measuring a flux, but with caution
  \citep[see][]{magnier2017.analysis,waters2017}.
\item {\bf rank 5 :} Photometric calibration of the GPC1 exposure is
  determined by relphot to be poor.  This situation occurs if there
  are too few stars available for the calibration ($< 10$ selected
  stars, or if the selected stars account for $< 5\%$ of all stars in
  the exposure).  An exposure may also be identified as poor if the
  zero point is excessively deviant ($> 2$ magnitudes from the nominal
  value) or if the standard deviation of the calibration residuals is
  more than $2\times$ the median standard deviation for all exposures.
\item {\bf rank 6 :} Photometry analysis flag field (\code{photFlags}) has one of the ``bad quality'' bits raised.  These bits are listed below; OR-ed together they have the hexadecimal value \code{0x1003bc88}
\begin{itemize}
  \item {\tt PM\_SOURCE\_MODE\_FAIL = 0x00000008} : Non-linear fit failed (non-converge, off-edge, run to zero)
  \item {\tt PM\_SOURCE\_MODE\_SATSTAR = 0x00000080} : Source model peak is above saturation
  \item {\tt PM\_SOURCE\_MODE\_BADPSF = 0x00000400} : Failed to get good estimate of object's PSF
  \item {\tt PM\_SOURCE\_MODE\_DEFECT = 0x00000800} : Source is thought to be a defect
  \item {\tt PM\_SOURCE\_MODE\_SATURATED = 0x00001000} : Source is thought to be saturated pixels (bleed trail)
  \item {\tt PM\_SOURCE\_MODE\_CR\_LIMIT = 0x00002000} : Source has crNsigma above limit
  \item {\tt PM\_SOURCE\_MODE\_MOMENTS\_FAILURE = 0x00008000} : Could not measure the moments
  \item {\tt PM\_SOURCE\_MODE\_SKY\_FAILURE = 0x00010000} : Could not measure the local sky
  \item {\tt PM\_SOURCE\_MODE\_SKYVAR\_FAILURE = 0x00020000} : Could not measure the local sky variance
  \item {\tt PM\_SOURCE\_MODE\_SIZE\_SKIPPED = 0x10000000} : Size could not be determined
\end{itemize}
\item {\bf rank 7 :} Measurement is from an invalid time period or
  photometry code.  This rank level is not used in the $3\pi$ PV3
  calibration.  Measurements were not restricted on the basis of the
  time of the observation, and only GPC1 measurements were explicitly
  included.
\item {\bf rank 8 :} Instrumental magnitude out of range.  This rank level was not used in the $3\pi$ PV3 calibration.
\end{itemize}

Rank values are assigned exclusively starting from the highest values:
if a measurements satisfies the rule for \eg, rank 6, it will not be
tested for ranks 5 and lower.  After all measurements have been
assigned a ranking value, the set of all measurements with the common
lowest value are selected to be used for the average photometry
analysis.  If measurements from ranks 0 through 4 were used for the
average photometry for a given filter, a per-filter mask bit value is
raised identifying which rank was used.  These bit are called
\code{ID_SECF_RANK_0} through \code{ID_SECF_RANK_4} (see
Table~\ref{tab:secf_mask_values}).  


\subsubsection{Iteratively Reweighted Least Squares Fitting}

With an automatic process applied to hundreds of millions of stars, it
is important for the analysis to provide a measurement of the
photometry of each object which is robust against failures.  The
Pan-STARRS\,1 detections have a relatively high rate of non-Gaussian
outliers, partly because of the wide range of instrumental features
affecting the data \citep[see][]{waters2017}.  We have used a
technique called Iteratively Reweighted Least Squares (IRLS) fitting
to reduce the sensitivity of the fits to outlier measurements.  We
have also used bootstrap resampling to determine confidence limits on
our fits given the observed collection of photometry measurements.  In
this case, the analysis is fitting the trivial model that the
photometry measurements are derived from a population with an
underlying constant value.  The discussion below applies to both the
average of the chip photometry magnitudes and the forced-warp
photometry fluxes.

The IRLS analysis starts with an ordinary least squares fit, using the
weights for each measurement as determined from Poisson statistics.
Since our model is a constant flux, this step is equivalent to
calculating a simple weighted average.  

Next, the deviations from the average value for each photometry
measurement are calculated.  The deviation, normalized by the Poisson
error, is used to modify the standard weight.  We use a Cauchy
function to define a new weight:
\begin{equation}
\omega^\prime = \frac{\omega}{1 + r^2}
\end{equation}
using
\begin{equation}
r = \frac{F_o - F_i}{\sigma}
\end{equation}
where $F_o$ is the average magnitude (or flux for forced-warp
photometry), $F_i$ is the measured magnitude (or flux), $\sigma$ is
the standard Poisson-based error on the photometry measurement, and
$\omega$ is the ordinary Poisson weight ($\sigma^{-2}$).  This
modified weight has the behavior that if the observed photometry
differs from the model by a substantial amount, the weight is greatly
reduced, while the weight approaches the standard weight if the model
and observed positions agree well.  Thus, this procedure is equivalent
to sigma clipping, but allows the outliers to be reduced in impact in
a continuous way, rather than rigidly accepting or rejecting them.

The weighted average photometry is re-calculated with these modified
weights.  New values for $\omega$ are calculated, and the weighted
average is calculated again.  On each iteration, the weighted average
photometry values are compared to the values from the previous
iteration.  If they have not changed significantly ($< 10^{-6}$) or if
the fractional change is less than some tolerance ($10^{-4}$), then
iterations are halted and the last weighted average values are used.
If convergence is not reached in 10 iterations, the process is halted
in any case and a flag raised for the object to note that IRLS did not
converge.


To calculate a fit $\chi^2$ value and to determine an appropriate set
of errors for the model parameters, it is necessary to transform the
modified weights into explicit cuts.  We have used the rubric that if
the modified weight is less than 30\% of the median weight
($\omega^\prime < 0.3 <\omega>$) then the point is treated as clipped.
The $\chi^2$ is determined from the {\em unclipped} points using the
standard Poisson errors.

Bootstrap-resampling analysis is used to assess the errors on the fit
parameters: A number of measurements equal to the number of {\em
  unclipped} data points are randomly selected from the set of
unclipped data points, with replacement after each selection.  These
data points are then used to calculate the weighted average
photometry.  The average values is recorded and the process re-run 100
times.  The error on the photometry value is determined as half of the
68\% confidence range for the distribution of average values.
However, if the number of measurements is small, the
bootstrap-resampled measurement of the error may be artificially
small.  We record the maximum of the bootstrap-sampling error and the
formal error from the weighted average calculation.  The minimum and
maximum of the unclipped values are also recorded for the chip
photometry.




\subsubsection{Stack Photometry}

For the stack photometry, the assessment is different from the chip
and forced-warp photometry: multiple measurements are not used to
calculate an average value.  For most of the sky, only a single set of
stack pixels exist for each filter.  Ideally, a unique astronomical
object would only be detected once in a given filter, resulting in
only a single measurement of that object from that filter's stack in
the database.  In practice, objects within a single stack image are
occasionally split by the analysis code, resulting in multiple
detections of the same object.  This situation is discussed in more
detail below.  

\begin{figure*}[htbp]
  \begin{center}
 \includegraphics[width=\hsize,clip]{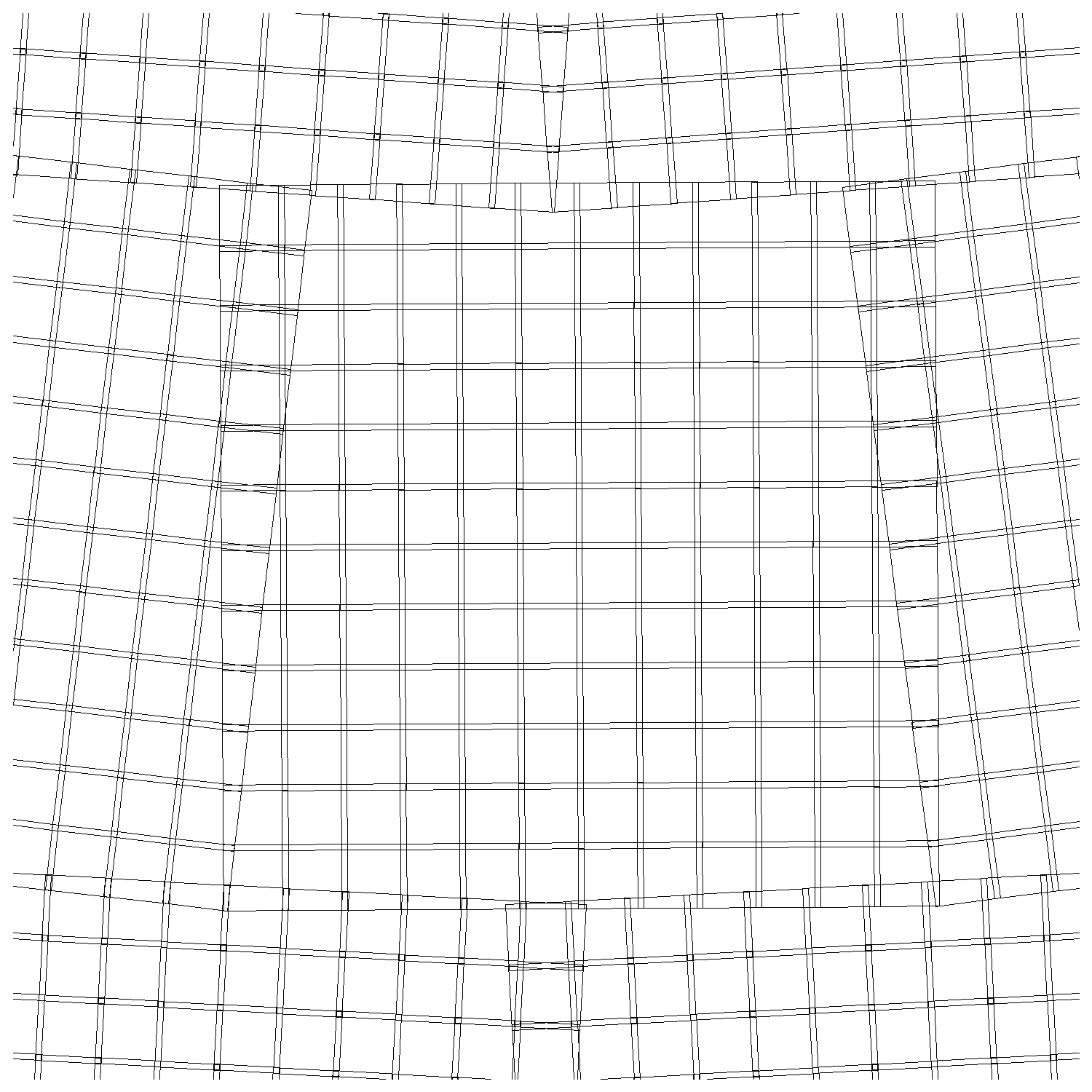}
  \caption{\label{fig:rings.v3.example} Illustration of overlapping
    skycells and the identification of the ``primary'' detections.}
  \end{center}
\end{figure*}

In addition to the these relatively rare failure cases, the objects
detected in the stacks may also have multiple measurements
due to the overlap between neighboring stack images.  The skycells 
(within which the stacks are generated) for a given projection cell
are defined to have significant overlap between neighbors to ensure that a
modestly-extended object can be measured completely on the pixels in a
single skycell image.  For the \ippmisc{RINGS.V3} skycell tessellation
used for the $3\pi$ PV3 analysis, this overlap was set to be 60
arcseconds, \ie, 240 extra pixels on each edge.  Within
\ippmisc{RINGS.V3}, projection cells themselves are defined to have an
overlap with neighboring projection cells to avoid gaps due to the
process of tiling the spherical sky with a series of flat
projections.  Due to the curved surface of the sky, the amount of
overlap between projection cells increases away from the celestial
equator.  Figure~\ref{fig:rings.v3.example} illustrates both skycell
and projection cell overlaps.

Overlapping stack regions are not statistically independent.  In the
typical circumstance, the same raw chip images are used to generate
the input warp images for the skycell on either side of the overlap.
Except for rare edge cases (\eg, an input warp which was rejected from
the stack for one side but not the other), exactly the same input raw
chip pixels contribute to all sets of stack pixels which overlap.  It
would therefore be statistically inappropriate to average the multiple
stack measurements from different overlapping skycells.  Instead, we
identify a unique set of stack measurements for the end user.

We identify two different ways in which an appropriate set of unique
stack measurements can be selected.  In the first case, if multiple
overlapping skycells contribute measurements to an object, we choose
the representative measurement based on their location in the skycell.
This selection is purely a function of the geometry of the skycells
and the coordinate of the object.  We first identify the primary
projection cells, those for which the overlapping regions are closest
to the projection cell center.  For regions in the primary projection
cell, we then identify the primary skycells, those for which the
overlapping regions are closest to the center of the skycell.  For a
given object, the identification of the primary projection cell and
skycell is calculated based on that the coordinates of the object.  We
then find the measurements for the object which came from the primary
projection cell and skycell and identify this set of measurements
(\grizy) as the ``primary'' set.  Note that we use the average
position of the object to define the ``primary'' measurements, forcing
measurements from all filters for the same skycell to be ``primary''
measurements, even if small deviations in the stack positions would
result in one of the filter detections falling on the other side of
the skycell ``primary'' boundary.  Thus, for a given object in the
database, we expect all 5 filters to provide a ``primary'' measurement
from the same skycell for each object.  Also note that a faint object,
near the detection limit of the stack, may be detected on a
secondary skycell but not (due to statistical fluctuations) be detected
on the corresponding primary skycell.  Thus it is expected that some
objects may be lacking any primary detections.

Since the ``primary'' identification is purely based on the skycell
geometry and the coordinate of the object, there is no guarantee that
any primary measurement is in fact a good or even the best measurement
of the object.  While the different overlapping pixels should be
essentially identical, it is possible (due to some of the edge cases
mentioned above) that one of the two sets of pixels is more heavily
masked than the other (\eg., more rejected inputs to the stack).
Thus, it is possible that one of the measurements is valid while the
other is not.  To address this possibility, we also identify a set of
``best'' measurements for each object.

For the stack measurements of an object in a specific filter, if there
are ``primary'' measurements with finite signal-to-noise and PSF
``perfect pixel'' quality factor (\code{PSF_QF_PERFECT}) $> 0.95$, the
measurement with the highest signal-to-noise is marked as ``best''.
If no primary measurement has \code{PSF_QF_PERFECT} $> 0.95$, but a
secondary measurement does, then the secondary measurement with the
highest signal-to-noise is chosen as ``best''.  If neither of the
first two cases hold, but there exist primary measurements with lower
\code{PSF_QF_PERFECT} values, the measurement with the highest
\code{PSF_QF_PERFECT} value is chosen as ``best''.  Finally, if no
``best'' value has yet been identified, the secondary measurement with
the highest value of \code{PSF_QF_PERFECT} is chosen as ``best''.
Note that the above rules allow for multiple measurements of the same
object from the same skycell pixels.  This may occur if the object was
split due to, \eg, saturation or complex morphology.  This type of
split should not be common (and in fact reflects a failure of the
algorithm), but we have defined the rules to allows us to choose an
acceptable measurement even in these cases.

\subsubsection{Warp Photometry}

The calculation of the average forced-warp photometry is performed
very similarly to the average of the chip photometry, with two
important exceptions.  First, as discussed above, the forced-warp {\em
  fluxes} are averaged, rather than the magnitudes.  Second, only the
warp measurements from the skycell which provided the ``best'' stack
measurements are used to calculate the average.  Just as the
overlapping stack pixels are not statistically independent,
overlapping warp pixels from the same exposure are also not
statistically independent. It is critical to use only a single
measurement from each input exposure.  We choose to use those from the
``best'' stack skycell rather than the ``primary'' stack skycell to
ensure the forced-warp photometry represents the highest quality set
of measurements.  Once the measurements from the chosen skycell have
been selected, the same quality cuts are applied to the measurements
as are applied to the chip measurements, as discussed above.

\begin{figure*}[htbp]
  \begin{center}
 \includegraphics[width=\hsize,clip]{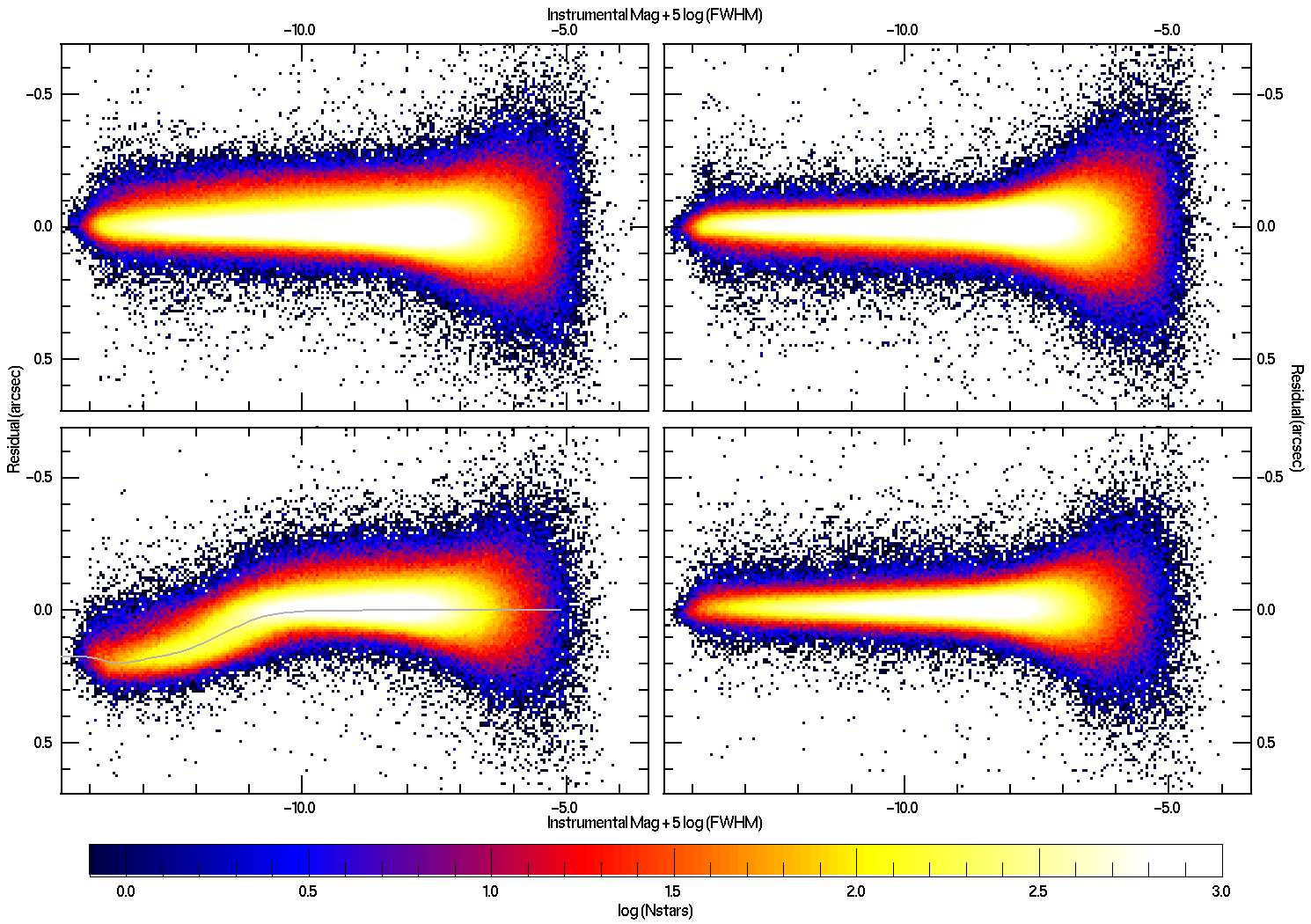}
  \caption{\label{fig:KHexample} Illustration of the Koppenh\"ofer Effect
    on chip XY04.  {\bf Bottom left} X-direction before correction.  The solid line shows the measured
    mean residual for stars detected on this chip as a function of the
    instrumental magnitude / FWHM$^2$.  
{\bf Bottom right} Y-direction before correction.  
{\bf Top left} X-direction after correction.  
{\bf Top right} Y-direction after correction.  }
  \end{center}
\end{figure*}

\begin{figure}[htbp]
  \begin{center}
 \includegraphics[width=\hsize,clip]{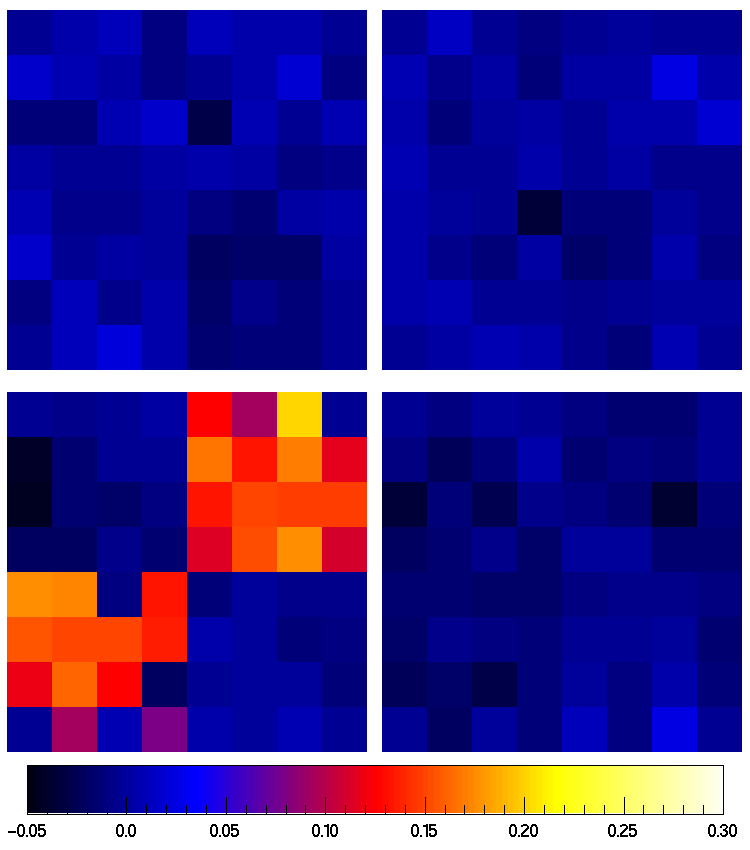}
  \caption{\label{fig:KHmap} Map of the amplitude of the
    Koppenh\"ofer Effect on chips across the focal plane.  In the
    affected chips, bright stars are up to 0.2 arcsec deviant
    from their expected positions. {\bf Bottom left} X-direction before
    correction.  {\bf Bottom right} Y-direction before correction.  {\bf
      Top left} X-direction after correction.  {\bf Top right}
    Y-direction after correction.}
  \end{center}
\end{figure}

\section{Astrometry Calibration}

Once the full PV3 dataset loaded into the master PV3 DVO database,
along with supporting databases, and the photometric calibrations were
performed, relative astrometry could be performed on the database to
improve the overall astrometric calibration.

In many respects the relative astrometric analysis is similar to the
relative photometric analysis: the repeated measurements of the same
object in different images are used to determine a high quality
average position for the object.  The new average positions are then
used to determine improved astrometric calibrations for each of the
images.  These improved calibrations are used to set the observed
coordinates of the measurements from those images, which are in turn
used to improve the average positions of the objects.  The whole
process is repeated for several iterations.  Like the photometric
analysis, the astrometric analysis is performed in a parallel fashion
with the same concept that specific machines are responsible for
exposures and objects which land within their regions of
responsibility, defined on the basis of lines of constant RA and DEC.
Between iteration steps, the astrometric calibrations are shared
between the parallel machines as are the improved positions for
objects controlled by one machine but detect in images controlled by
another machine.  Like the photometric analysis, the entire sky is
processed in one pass.  However, there are some important differences
in the details.

\subsection{Systematic Effects}

First, the astrometric calibration has a larger number of systematic
effects which must be performed.  These consist of: 1) the
Koppenh\"ofer Effect, 2) Differential Chromatic Refraction, 3) Static
deviations in the camera.  We discuss each of these in turn below.

\subsubsection{Koppenh\"ofer Effect}

The Koppenh\"ofer Effect was first identified in February 2011 by
Johannes Koppenh\"ofer (MPE) as part of the effort to search for
planet transits in the Stellar Transit Survey data.  He noticed that
the astrometry of bright stars and faint stars disagreed on overlapping
chips at the boundary between the STS fields.  After some exploration,
it was determined that the X coordinate of the brightest stars was
offset from the expected location based on the faint stars for a
subset of the GPC1 chips.  The essence of the effect was that a large
charge packet could be drawn prematurely over an intervening negative
serial phase into the summing well, and this leakage was
proportionately worse for brighter stars.  The brighter the star, the
more the charge packet was pushed ahead on the serial register.  The
amplitude of the effect was at most $0\farcs{}25$, corresponding to a
shift of about one pixel.  This effect was only observed in 2-phase
OTA devices, with 22 / 30 of these suffering from this effect.  By
adjusting the summing well high voltage down from a default +7 V to
+5.5V on the 2-phase devices, the effect was prevented in exposures
after 2011-05-03.  However, this left 101,550 exposures (27\%) already
contaminated by the effect.


We measured the Koppenh\"ofer Effect by accumulating the residual
astrometry statistics for stars in the database.  For each chip, we
measured the mean X and Y displacements of the astrometric residuals
as function of the instrumental magnitude of the star divided by the
FWHM$^2$.  We measured the trend for all chips in a
number of different time ranges and found the effect to be quite
stable, in the period where it was present.  The effect only appeared
in the serial direction.  Figure~\ref{fig:KHexample} shows the KE
trend for a typical affected chip both before and after the
correction.  For the PV3 dataset, we re-measured the KE trends using
stars in the Galactic pole regions after an initial relative
astrometry calibration pass: the Galactic pole is necessary because
the real-time astrometric calibration relies largely on the fainter
stars which are not affected by the KE.  The trend is then stored in a
form which can be applied to the database measurements.

\subsubsection{Differential Chromatic Refraction}

Differential Chromatic Refraction (DCR) affects astrometry because the
reference stars used to the calibrate the images are not the same
color (SED) as the rest of the stars in the image.  For a given star
of a color different from the reference stars, as exposures are taken
at higher airmass, the apparent position of the star will be biased
along the parallactic angle.  While it is possible to build a model
for the DCR impact based on the filter response functions and
atmospheric refraction, we have instead elected to use an empirical
correction for the DCR present in the PV3 database.  We have measured
the DCR trend using the astrometric residuals of millions of stars
after performing an initial relative astrometry calibration.  We
define a blue DCR color ($g-i$) to be used when correcting the filters
\gps,\rps,\ips, and a red DCR color ($z - y$) to be used when
correcting the filters $zy$.  In the process of performing the
relative astrometry calibration, we record the median red and blue
colors of the reference stars used to measure the astrometry
calibration for each image.  As we determine the astrometry parameters
for each object in the database, we record the median red and blue
reference star colors for all images used to determine the astrometry
for a given object.  For each star in the database, we know both the
color of the star and the typical color of the reference stars used to
calibrate the astrometry for that star.

\begin{figure}[htbp]
  \begin{center}
 \includegraphics[width=\hsize,clip]{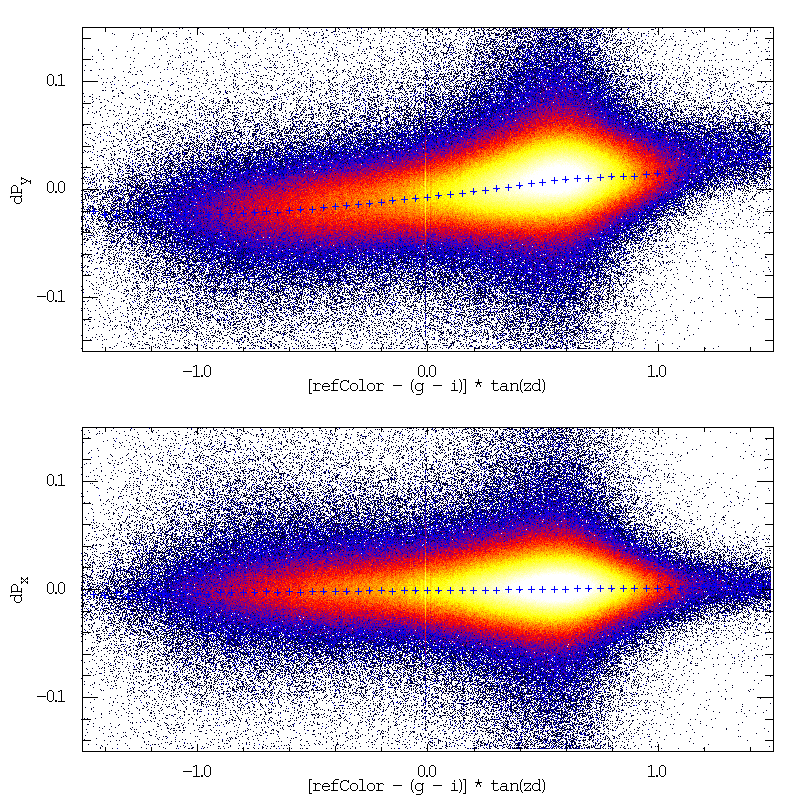}
  \caption{\label{fig:DCRexample} Example of the DCR trend in the
    g-band.  {\bf top:} DCR trend in the parallactic direction {\bf
      bottom:} DCR trend perpendicular to the parallactic angle.}
  \end{center}
\end{figure}

We measure the mean deviation of the residuals in the parallactic
angle direction and the direction perpendicular to the parallactic
angle.  For each filter, we determine the DCR trend as a function of
the difference between the star color and the reference star color,
using the red or blue color appropriate to the particular filter, times
the tangent of the zenith distance.  Figure~\ref{fig:DCRexample} shows the
DCR trend for the 5 filters \grizy, as well as the measured
displacement in the direction perpendicular to the parallactic angle.
We represent the trend with a spline fitted to this dataset.  

The amplitude of the DCR trend in the five filters is $(g,r,i,z,y) =
(0.010, 0.001, -0.003, -0.017, -0.021)$ arcsec airmass$^{-1}$
magnitude$^{-1}$.  We saturate the DCR correction if the term $color
TAN (\zeta)$ for a given measurement is outside a range where the
DCR correction is well measured.  The maximum DCR correction applied
to the five filters is $(g,r,i,z,y) = (0.019, 0.002, 0.003, 0.006,
0.008)$ arcseconds.


\begin{figure*}[htbp]
 \begin{center}
 \includegraphics[width=0.85\textwidth,clip]{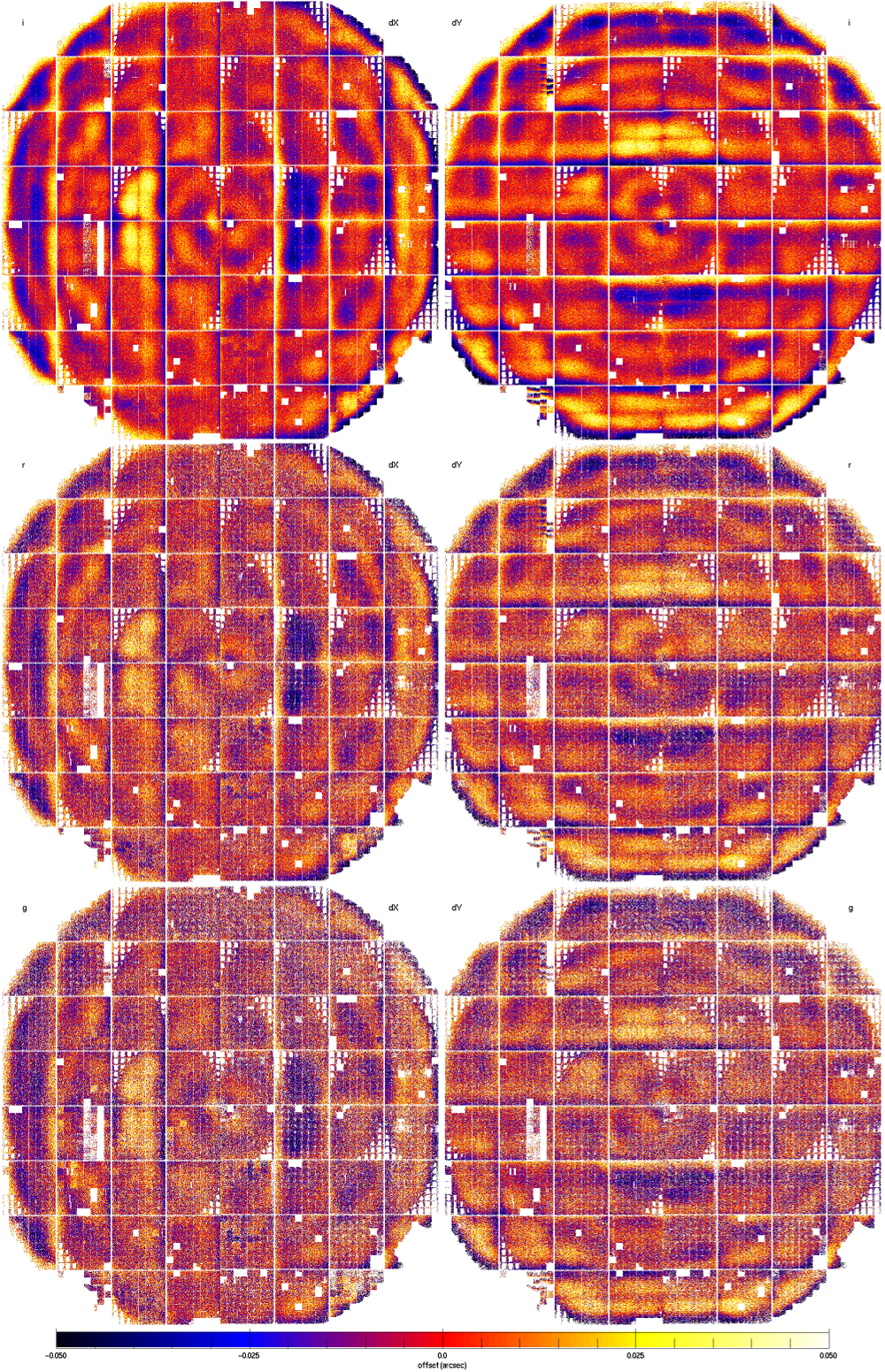}
 \caption{\label{fig:astroflat.gri} High-resolution astrometric flat-field correction images for $gri$.}
 \end{center}
\end{figure*}

\begin{figure*}[htbp]
 \begin{center}
 \includegraphics[width=0.85\textwidth,clip]{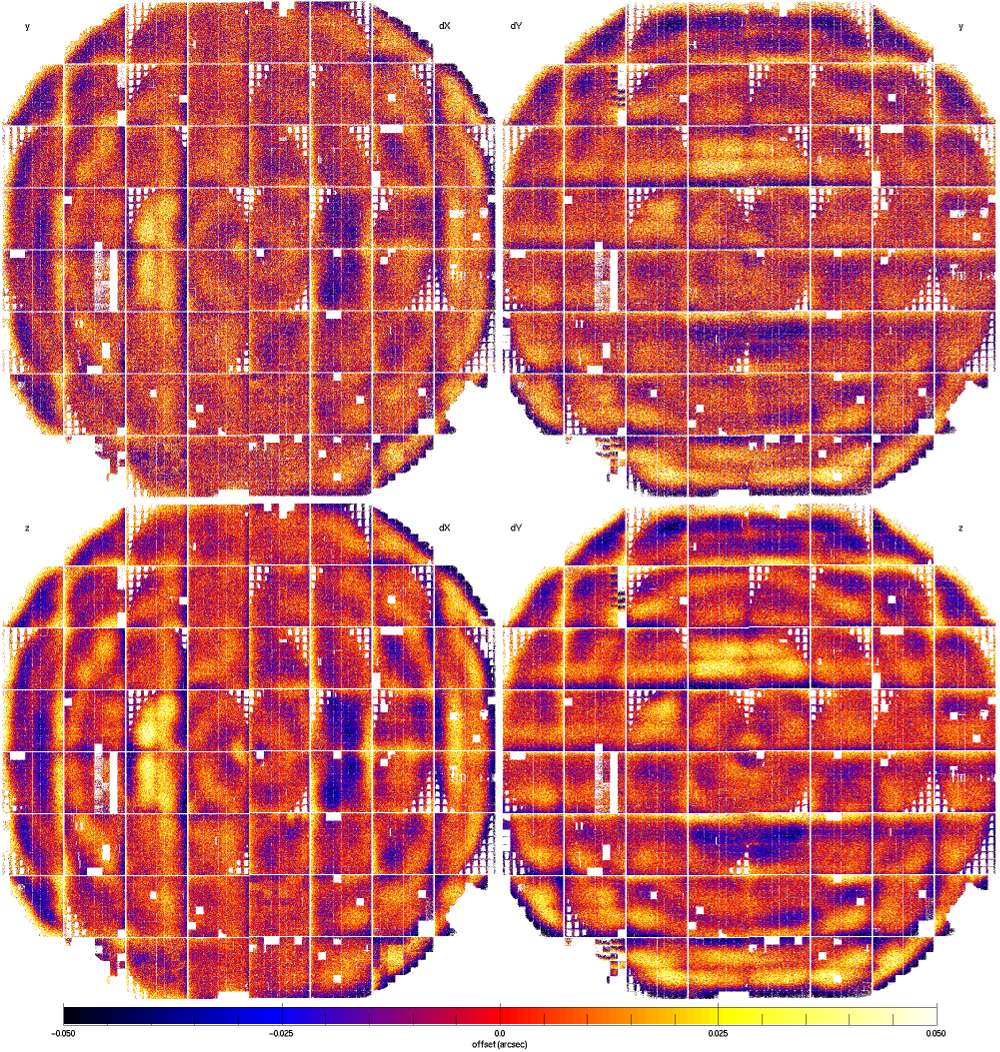}
 \caption{\label{fig:astroflat.zy} High-resolution astrometric flat-field correction images for $zy$.}
 \end{center}
\end{figure*}

\subsubsection{Astrometric Flat-field}

After correction for both KE and DCR, we observe persistent residual
astrometric deviations which depend on the position in the camera.  We
construct an astrometric ``flat-field'' response by determining the
mean residual displacement in the X and Y (chip) directions as a
function of position in the focal plane.  We have measured the
astrometric flat using a sampling resolution of 40x40 pixels, matching
the photometric flat-field correction images.
Figures~\ref{fig:astroflat.gri} and \ref{fig:astroflat.zy} show the
astrometric flat-field images for the five filters \grizy\ in each of
the two coordinate directions.  These plots show several types of
features.


The dominant pattern in the astrometric residual is roughly a series
of concentric rings. The pattern is similar to the pattern of the
focal surface residuals measured by \cite{2008SPIE.7014E..0DO}, which also has
a concentric series of rings with similar spacing.  The ``tent'' in
the center of the focal surface is reflected in these astrometry
residual plots.  Our interpretation of the structure is that the
deviations of the focal plane from the ideal focal surface introduces
small-scale PSF changes, presumably coupled to the optical
aberrations, which result in small changes in the centroid of the
object relative to the PSF model at that location.  Since the PSF
model shape parameters are only able to vary at the level of a 6x6
grid per chips, the finer structures are not included in the PSF
model.  The PV2 analysis shows the ring structure more clearly, with a
pattern much more closely following the focal surface deviations.  In
the PV2 analysis, the PSF model used at most a 3x3 grid per chip to
follow the shape variations, so any changes caused by the optical
aberrations would be less well modeled in the PV2 analysis, as we
observe.

A second pattern which is weakly seen in several chips consists of
consistent displacements in the X (serial) direction for certain
cells.  This effect can be seen most clearly in chips XY45 and XY46.
In the PV2 analysis, this pattern is also more clearly seen.  In this
case, the fact that the astrometric model used polynomials with a
maximum of 3rd order per chip means the deviation of individual cells
cannot be followed by the astrometric model.  

A third effect is seen at the edge of the chips, where there appears
to be a tendency for the residual to follow the chip edge.  The origin
of this is unclear, but likely caused by the astrometry model failing
to follow the underlying variations because of the need to extrapolate
to the edge pixels.  Finally, we also mention an interesting effect
{\em not} visible at the resolution of these astrometric flat-field
images.  Fine structures are observed at the \approx 10 pixel scale
similar to the ``tree rings'' reported by the Dark Energy Survey team
\citep{2014PASP..126..750P} and identified as a result of lateral
diffusion of electrons in the detectors due to electric fields due to
dopant variations.  Unlike the photometric tree ring features
discussed above (Section~\ref{sec:phot.flat}), these astrometric tree
rings appear to correspond to the features identified by the DES team.
Lateral electric fields in the detector silicon, caused by variations
in the dopant density, cause the photoelectrons to migrate laterally
in the detector silicon before landing in the pixel wells.  This
migration affects the apparent position of the stars, thus affecting
the observed astrometry.  A simple lateral translation of the
effective pixel locations would not be detected as it would be
degenerate with the astrometric solution.  However, since the lateral
electric fields, and thus the electron migration, vary with position,
the astrometric displacement changes on small scales relative to the
average solution, resulting in residual astrometric structures.  The
gradient of the astrometric displacement results in an apparent
expansion or compression of the pixel sizes, resulting in a signal
which can be observed in the flat-field images.  For GPC1, unlike the
DES detectors, the amplitude of these flat-field variations are much
smaller than the photometric variations caused by the changing PSF
sized, caused in turn by varying electron diffusion rates.  These
features, and the related vertical electron diffusion variations are
discussed in detail in \cite{2018PASP..130f5002M}.

Unfortunately, we discovered a problem with the astrometric flat-field
correction too late to be repaired for DR1.  As can be seen by
inspection of Figures~\ref{fig:astroflat.gri} and
\ref{fig:astroflat.zy}, there is significant pixel-to-pixel noise in
the the astrometric flat-field images.  This pixel-to-pixel noise is
caused by too few stars used in the measurement of the flat-field
structure for the high-resolution sampling.  As a result, the
astrometric flat-field correction reduces systematic structures on
large spatial scales, but at the expense of degrading the quality of
an individual measurement.  Only $i$-band has sufficient
signal-to-noise per pixel to avoid significantly increasing the
per-measurement position errors.  

\begin{figure}[htbp]
  \begin{center}
 \includegraphics[width=\hsize,clip]{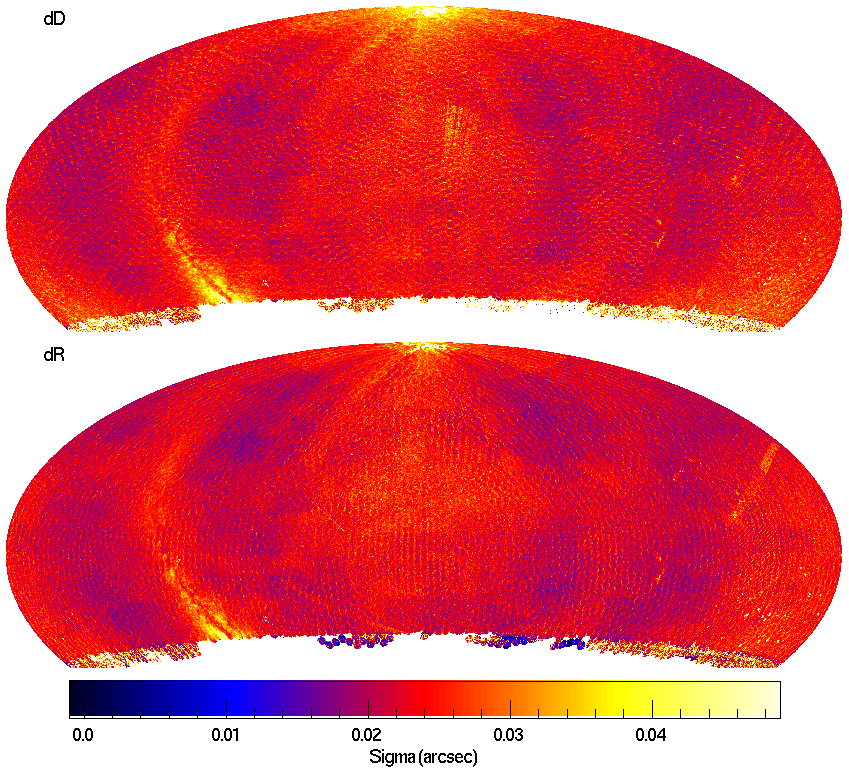}
  \caption{\label{fig:allsky.astrom.sigma} Consistency of photometry
    measurements across the sky.  Each panel shows a map of the
    standard deviation of astrometry residuals for stars in each
    pixel.  The median value of the standard deviations across the sky
    is $(\sigma_\alpha, \sigma_\delta) = (22, 23)$ milliarcseconds.
    These values reflect the typical single-measurement errors for
    bright stars.  See discussion regarding the astrometric flat which
    is likely responsible for these elevated value. }
  \end{center}
\end{figure}

Figure~\ref{fig:allsky.astrom.sigma} shows the standard deviations of
the mean residual astrometry in $(\alpha,\delta)$ for bright stars as
a function of position across the sky.  For each pixel in these
images, we selected all objects with $15 < i < 17$, with at least 3
measurements in $i$-band (to reject artifacts detected in a pair of
exposures from the same night), with \code{PSF_QF} $> 0.85$ (to reject
excessively-masked objects), and with $mag_{\rm PSF} - mag_{\rm Kron}
< 0.1$ (to reject galaxies).  We then generated histograms of the
difference between the object position predicted for the epoch of each
measurement (based on the proper motion and parallax fit) and the
observed position of that measurement, in both the Right Ascension and
Declination directions (in linear arcseconds), for all stars in a
given pixel in the images.  From these residual histograms, we can
then determine the median and the 68\%-ile range to calculate a robust
version of the standard deviation.  This represents the bright-end
systematic error floor for a measurement from a single exposure.  The
standard deviations are then plotted in
Figure~\ref{fig:allsky.photom.sigma}.  The median value of the
standard deviations across the sky is $(\sigma_\alpha, \sigma_\delta)
= (22, 23)$ milliarcseconds.

The Galactic plane is clearly apparently in these images.  Like
photometry, we attribute this to failure of the PSF fitting due to
crowding.  The celestial North pole regions have somewhat elevated
errors in both R.A. and DEC.  This may be due to the larger typical
seeing at these high airmass regions, but without further exploration
this interpretation is uncertain.  Several features can be seen which
appear to be an effect of the tie to the Gaia astrometry: the stripes
near the center of the DEC image and the right side of the R.A. image.
The mesh of circular outlines is due to the outer edge of the focal
plane where the astrometric calibration is poorly determined.  As
discussed above, the median values in the images are higher than
expected based on our PV2 analysis of the astrometry: the median
per-measurement error floor of \approx 22 mas is significantly worse
than the \approx 17 mas value in that earlier analysis.  We attribute
this degradation to the noise introduced by the astrometric
flat-field.  This noise has been addressed for the DR2 release
of the individual measurement data.

After the initial analysis to measure the KE corrections, DCR
corrections, and astrometric flat-field corrections, we applied these
corrections to the entire database.  Within the schema of the
database, each measurement has the raw chip coordinates
(\code{Measure.Xccd,Yccd}) as well as the offset for that object based on each of
these three corrections: \code{Measure.XoffKH,YoffKH,
  Measure.XoffDCR,YoffDCR, Measure.XoffCAM,YoffCAM}.  The offsets are
calculated for each measurement based on the observed instrumental
chip magnitudes and FWHM for the Koppenh\"ofer Effect, on the average
chip colors and the altitude \& azimuth of each measurement for the
DCR correction, and on the chip coordinates for the astrometric
flat-field corrections.  The corrections are combined and applied to
the raw chip coordinates and saved back in the database in the fields
\code{Measure.Xfix,Yfix}.  At this point, we are ready to run the
full astrometric calibration. 

\subsection{Galactic Rotation and Solar Motion}

The initial analysis of the PV2 astrometry used the 2MASS positions as
an inertial constraint: the 2MASS coordinates were included in the
calculation of the mean positions for the objects in the database,
with weight corresponding to the reported astrometric errors.  In this
analysis, the object positions used to determine the calibrations of
the image parameters ignored proper motion and parallax.  After the
image calibrations were determined, then individual objects were
fitted for proper motion and possibly parallax, as discussed in detail
below. 

Using the PV2 analysis of the astrometry calibration, we discovered
large-scale systematic trends in the reported proper motions of
background quasars.  This motion had an amplitude of 10 - 15
milliarcseconds per year and clear trends with Galactic longitude.  We
also observed systematic errors of the mean positions with respect to
the ICRF milliarcsecond radio quasar positions, with an amplitude of
\approx 60 milliarcseconds, again with trends associated with Galactic
longitude.  Since the 2MASS data were believed to have minimal average
deviations relative to the ICRF quasars, this latter seemed to be a
real effect.  

We realized that both the proper motion and the mean position biases
could be caused by a single common effect: the proper motion of the
stars used as reference stars between the 2MASS epoch (\approx 2000)
and PS1 epoch (\approx 2012).  Since we are fitting the image
calibrations without fitting for the proper motions of the stars, we
are in essence forcing those stars to have proper motions of 0.0.
The background quasars would then be observed to have proper motions
corresponding to the proper motions of the reference stars, but in the
opposite direction.  We demonstrated that the observed quasar proper
motions agreed well with the distribution expected if the median
distance to our reference stars was \approx 500 pc.  

For PV3, we desired to address this bias by including our knowledge
about the distances to the reference stars and the expected typical
proper motions for stars at those distances.  With some constraint on
the distance to each star, we can determine the expected proper motion
based on a model of the Galactic rotation and solar motions.  We can
then calculate the mean positions for the objects keeping the assumed
proper motion fixed.  When calibrating a specific image, the reference
star mean position is then translated to the expected position at the
epoch of that image.  The image calibration is then performed relative
to these predicted positions.  This process naturally accounts for the
proper motion of the reference stars.  In order to make the
calibrations consistent with the observed coordinates of an external
inertial reference, we perform the iterative fits using the technique
as described, but assign very high weights in the initial iterations
to the inertial reference, and reduce the weights as the astrometric
calibration iterations proceed.

In order to perform this analysis, we need estimated distances for
every reference star used in the analysis.  \cite{2014ApJ...783..114G}
performed SED fitting for 800M stars in the 3$\pi$ region using PV2
data.  The goal of this work was to determine the 3D structure of the
dust in the galaxy.  By fitting model SEDs to stars meeting a basic
data quality cut, they determined the best spectral type, and thus
$T_{\rm eff}$, absolute $r$-band magnitude, distance modulus, and
extinction $A_V$ (the desired output and used to determine the dust
extinction as a function of distance throughout the galaxy).  We use
the distance modulus determined in this analysis to predict the proper
motions.

To convert the distances to proper motions, we use the Galactic
rotation parameters ($A,B$) = (14.82,-12.37) km sec$^{-1}$ pc$^{-1}$
and Solar motion parameters ($U_{\rm sol}, V_{\rm sol}, W_{\rm sol}$)
= (9.32, 11.18, 7.61) km sec$^{-1}$ as determined by
\cite{1997MNRAS.291..683F} using Hipparchus data.  Proper motions are
determined from the following:
\begin{eqnarray}
\mu^{\rm gal}_{l} & = & (A \cos (2 l) + B) \cos (b) \\
\mu^{\rm gal}_{b} & = & \frac{-A \sin (2 l) \sin (2 b)}{2} \\
\mu^{\rm sol}_{l} & = & \frac{U \sin(l) - V \cos(l)}{d} \\
\mu^{\rm sol}_{b} & = & \frac{(U \cos(l) + V \sin(l)) \sin(b) - W \cos(b)}{d}
\end{eqnarray}
where $d$ is the distance and $l,b$ are the Galactic coordinates of the
star. Note that the proper motion induced by
the Galactic rotation is independent of distance while the reflex
motion induced by the solar motion decreases with increasing
distance.  Also note that this model assumes a flat rotation curve for
objects in the thin disk; any reference stars which are part of 
the halo population will have proper motions which are not 
described by this model; the mostly random nature of the halo motions
should act to increase the noise in the measurement, but should not
introduce detectable motion biases.  Also, if the distance modulus is
not well determined, we can assume the object is simply following the
Galactic rotation curve and set a fixed proper motion.  If we do not
have a distance modulus from the Green et al analysis, we assume a
value of 500pc.  


\subsection{Gaia Constraint}

After the full relative astrometry analysis was performed for the PV3
database, the Gaia Data Release 1 became available
\citep{2016AA...595A...2G,2016AA...595A...4L}.  This afforded us
the opportunity to constrain the astrometry on the basis of the Gaia
observations.  Gaia DR1 objects which are bright enough to have proper
motion and parallax solutions are in general saturated in the PS1
observations.  Thus, we are limited to using the Gaia mean positions
reported for the fainter stars.  We extracted all Gaia sources not
marked as a duplicate from the Gaia archive and generated a DVO
database from this dataset.  We then merged the Gaia DVO into the PV3
master DVO database.  We re-ran the complete relative astrometry
analysis using Gaia as an additional measurement.  We applied the
analysis described above, applying the estimated distances to
determine preliminary proper motions.  The Gaia mean epoch is reported
as 2015.0, so all Gaia measurements were assigned this epoch.  We
wanted to ensure the Gaia measurements dominated the astrometric
solutions, so we made the weight very high for the Gaia points:
1000$\times$ the nominal weight in the initial fits (to lock down the
reference frame), decreasing to 100$\times$ the nominal weight for the
last fits.  We also retained the 2MASS measurements in the analysis,
but gave them somewhat lower weights than Gaia: while the 2MASS data
does not have the accuracy of Gaia, the coverage is known to be quite
complete, while the Gaia DR1 has clear gaps and holes.  Having 2MASS,
even at a lower weight, helps to tile over those gaps.


\begin{figure*}[htbp]
  \begin{center}
  \includegraphics[width=\hsize,clip]{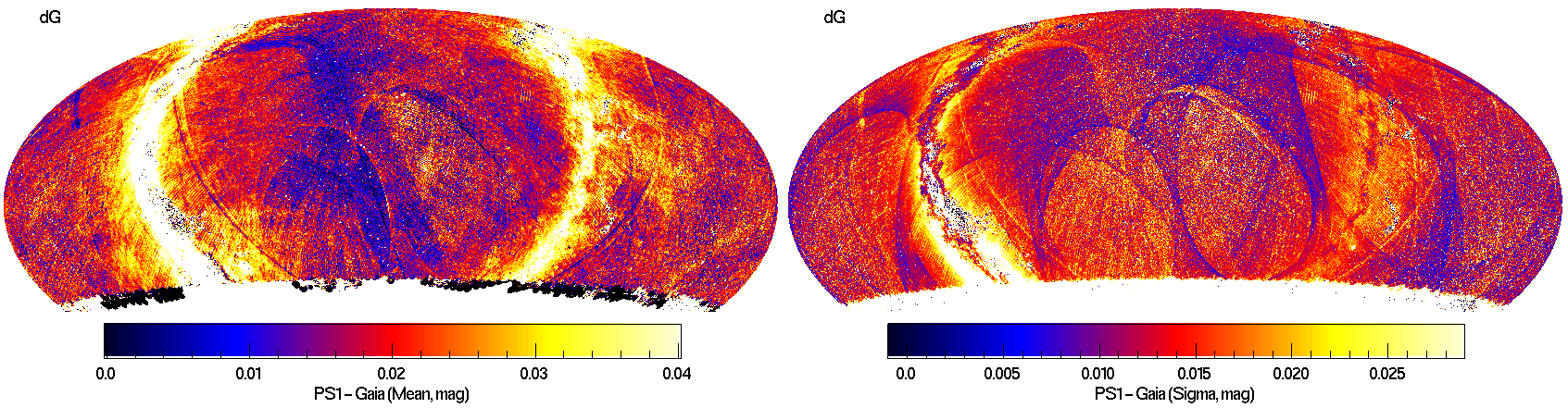}
  \caption{\label{fig:gaia.photom} Comparison with Gaia
    photometry. {\bf Left} Mean of PS1 - Gaia, {\bf Right} Standard
    deviation of PS1 - Gaia.  For pixels with $|b| > 30$ and $\delta >
    -30$, the standard deviation of the PS1 - Gaia mean values is 7
    millimagnitudes, while the median of the standard deviations is 12
    millimagnitudes.  The former is a statement about the consistency
    of the Gaia and Pan-STARRS\,1 photometry, while the latter
    reflects the combined bright-end errors for both systems.  }
  \end{center}
\end{figure*}

Figure~\ref{fig:gaia.photom} shows a comparison between the Pan-STARRS
photometry in $g,r,i$ and the Gaia photometry in the $G$-band.  To
compare the PS1 photometry to the very broadband Gaia G filter, we
have determined a transformation based on a 3rd order polynomial fit
to $g-r$ and $g-i$ colors.  This transformation reproduces Gaia
photometry reasonably well for stars which are not too red.  For a
comparison, we have selected all PS1 stars with Gaia measurements
meeting the following criteria: $14 < i < 19$, with at least 10 total
measurements, within a modest color range $0.2 < g - r < 0.9$.  We
also restricted to objects with $i_{\rm PSF} - i_{\rm Kron} < 0.1$,
using the average $i$ magnitudes determined from the individual
exposures.  

For Figure~\ref{fig:gaia.photom}, we calculate the difference between
the estimated $G$-band magnitude based on PS1 $g,r,i$ photometry and
the $G$-band photometry reported by Gaia.  For each pixel, we
determine the histogram of these differences and calculate the median
and the 68\%-ile range.  In Figure~\ref{fig:gaia.photom}, these
values are plotted as a color scale.  

The Galactic plane is clearly poorly matched between the two
photometry systems.  This may in part be due to the difficulty of
predicting $G$-band magnitudes for stars which are significantly
extincted: the $G$-band includes significant flux from the PS1
$z$-band which was not used in our transformation.  Many other large
scale feature in the median differences have structures similar to the
Gaia scanning pattern (large arcs and long parallel lines.  There are
also structures related to the PS1 exposure footprint.  These show up
as a mottling on the \approx 3 degree scale (e.g., lower right below
the Galactic plane).  The amplitude of the residual structures is
fairly modest.  The standard devition of the median difference values
is 7 millimagnitudes.  This number gives an indication of the overall
photometric consistency of both Gaia and PS1 and implies that the
systematic error floor for each survey is less than 7 millimags.



\begin{figure*}[htbp]
  \begin{center}
  \includegraphics[width=\hsize,clip]{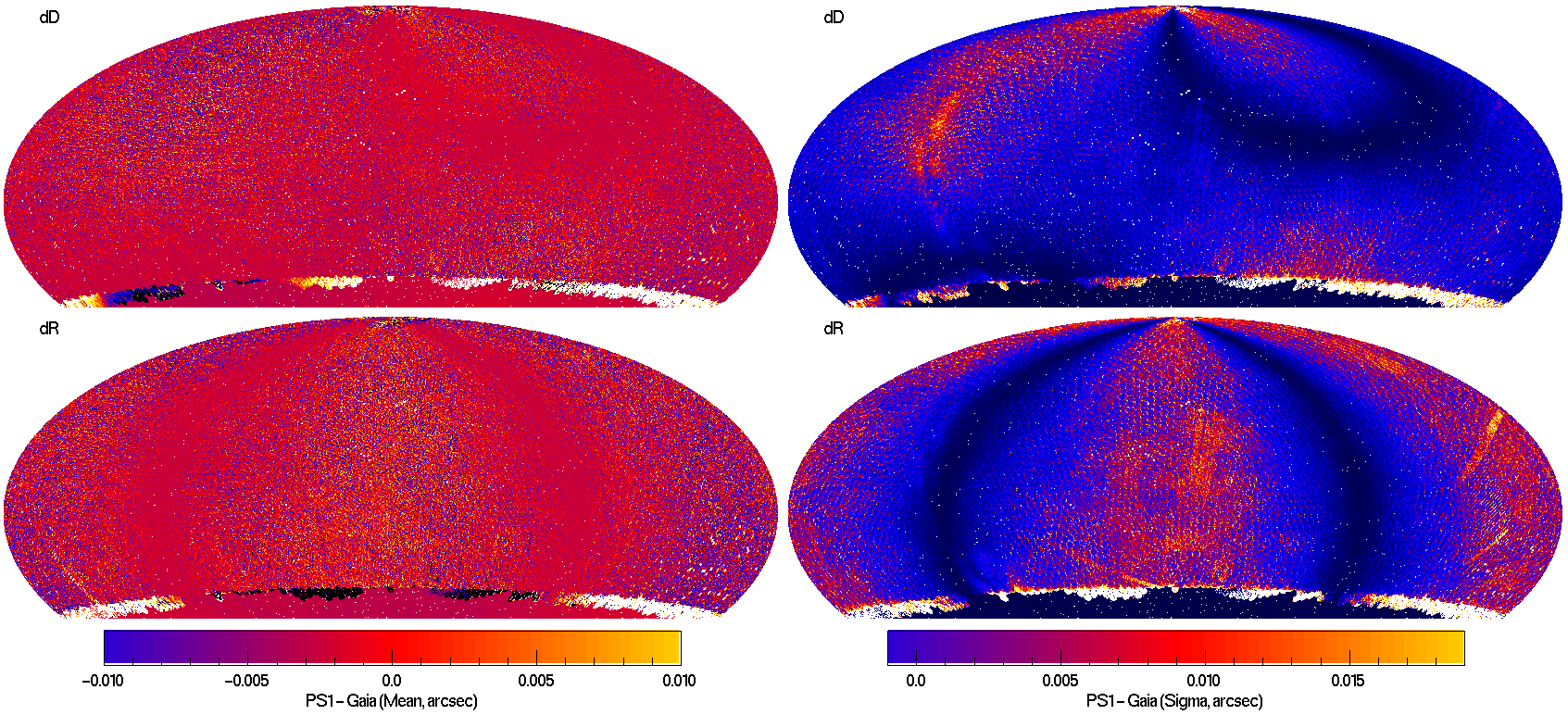}
  \caption{\label{fig:gaia.astrom} Comparison with Gaia
    astrometry. {\bf Left} Mean of PS1 - Gaia, {\bf Right} Standard
    deviation of PS1 - Gaia.  The median value of the standard
    deviations is $(\sigma_\alpha, \sigma_\delta) = (4, 3)$
    milliarcseconds. }
  \end{center}
\end{figure*}

Figure~\ref{fig:gaia.astrom} shows a comparison between the Pan-STARRS
mean astrometry positions in $\alpha,\delta$ and the Gaia astrometry.
For this comparison, we have seleted all PS1 stars with Gaia
measurements with $14 < i < 19$ and with at least 10 total
measurements.  For Figure~\ref{fig:gaia.astrom}, we calculate the
difference between the position predicted by PS1 at the Gaia epoch
(using the proper motion and parallax fit) and the position reported
by Gaia.  For each pixel, we determine the histogram of these
differences in the R.A\. and DEC directions, and calculate the median
and the 68\%-ile range.  In Figure~\ref{fig:gaia.astrom}, these
values are plotted as a color scale.

There is good consistency between the PS1 and Gaia astrometry.  There
are patterns from the Galactic plane (though not very strongly at the
bulge).  There are also clear features due to the PS1 exposure
footprint (ring structure on \approx 3 degree scales).  In the plots
of the scatter, there are patterns which are related to the Gaia
scanning rule.  These are presumably regions with relatively low
signal to noise in Gaia; they were also apparent in the plots of the
statisics of the per-exposure measurement residuals
(Figure~\ref{fig:allsky.astrom.sigma}.  The standard deviations of the
median differences are ($\sigma_\alpha, \sigma_\delta) = (4, 3)$
milliarcseconds.

For a future data release, we will recalibrate the Pan-STARRS $3\pi$
astrometry using the Gaia DR2 release.  The addition of Gaia-measured
proper motions will obviate the need to correct for the Galactic rotation.

\subsection{Calculation of Object Astrometry}

\subsubsection{Iteratively Reweighted Least Squares Fitting}

After the image astrometric parameters have been determined and
applied to the measurements from each image, we attempt to find
the best astrometric parameters (position, parallax and proper
motions) for all objects in the database.  We require a minimum of 5
detections and 1 year of data for any object in order for it to be
fitted for just proper motion.  For a parallax and proper-motion fit,
we require at least 7 detections, 1 year of data, and a parallax
factor range of at least 0.25; no object is fitted to parallax without
proper motion as well.  If an object is fitted for parallax, it is
also fitted with a model including only proper motion and only a mean
position.  The chisq for all three fits is saved.  Currently, the
highest order fit allowed is saved in the database, regardless of the
significance of the improvement in adding parameters.  The resulting
parallax and proper motion measurements are inserted back into the DVO
database for use by science queries.

With an automatic process applied to hundreds of millions of stars, it
is important for the analysis to provide a measurement of the
astrometry of each object which is robust against failures.  The
Pan-STARRS\,1 detections have a relatively high rate of non-Gaussian
outliers, partly because of the high degree of structure in the
astrometric transformations introduced by the camera optics and the
atmosphere, and partly due to the high masked fraction and other
detector effects.  We have used a techinique called Iteratively
Reweighted Least Squares (IRLS) fitting to reduce the sensitivity of
the fits to outlier measurements.  We have also used bootstrap
resampling to determine confidence limits on our fits given the
observed collection of position measurements.

We begin the astrometric analysis for each object by projecting the
sky coordinates ($\alpha,\delta$) to a locally linear coordinate
system ($\eta,\zeta$).  We choose as a reference a single measurement
from the full set of measurements.  It is not critical which
measurement we choose as long as the value is recorded during the
analysis so the results can be deprojected back to the sky using the
same reference coordinate.  We also work in a time system which has
been adjusted with reference to the average epoch from the collection
of measurements.  The resulting proper motions are thus determined
with the minimum degeneracy with respect to the average position
solution.

The IRLS analysis starts with an ordinary least squares fit, using the
weights for each measurement as determined from Poisson statistics.
After the astrometric parameters have been fitted, the deviations from
the fit for each position are calculated for both the local $\eta$ and
$\zeta$ coordinate directions.  The deviation, normalized by the
Poisson error, is used to modify the standard weight.  We use a Cauchy
function to define a new weight:
\begin{eqnarray}
\omega_\eta^\prime = \frac{\omega_\eta}{1 + r_\eta^2}\\
\omega_\zeta^\prime = \frac{\omega_\zeta}{1 + r_\zeta^2}\\
\end{eqnarray}
using
\begin{eqnarray}
r_\eta = \frac{\eta_o - \eta_i}{\sigma_\eta} \\
r_\zeta = \frac{\zeta_o - \zeta_i}{\sigma_\zeta}
\end{eqnarray}
where $\eta_o$ is the model position in the $\eta$ direction, $\eta_i$
is the measured position in the $\eta$ direction, $\sigma_\eta$ is the
standard error on the position in the $\eta$ direction, and
$\omega_\eta$ is the ordinary Poisson weight in the $\eta$ direction
($\sigma_\eta^{-2}$), and equivalently for the $\zeta$ direction.
This modified weight has the behavior that if the observed position
differs from the model by a substantial amount, the weight is greatly
reduced, while the weight approaches the standard weight if the model
and observed positions agree well.  Thus, this procedure is equivalent
to sigma clipping, but allows the outliers to be reduced in impact in
a continuous way, rather than rigidly accepting or rejecting them.

The object astrometric parameters are re-fitted with these modified
weights.  New values for $\omega_\eta,\omega_\zeta$ are calculated,
and the fit is tried again.  On each iteration, the fitted parameters
are compared to the values from the previous iteration.  If they
parameters have not changed significantly ($< 10^{-6}$) or if the
fractional change is less than some tolerance ($10^{-4}$), then
iterations are halted and the last fitted parameters are used.  If
convergence is not reached in 10 iterations, the process is halted in
any case and a flag raised for the object to note that IRLS did not
converge.


To calculate a fit $\chi^2$ value and to determine an appropriate set
of errors for the model parameters, it is necessary to transform the
modified weights into explicit cuts.  We have used the rubric that if
the modified weight is less than 30\% of the standard weight
($\omega^\prime_\eta < 0.3 \omega_\eta$) then the point is treated as
clipped.  If a data point would be clipped based on the modified
weight in either dimension, it is clipped in both (thus a point is
either used to calculate both RA and Declination terms, or neither).
The $\chi^2$ is determined from the unclipped points in the standard
way.  Bootstrap analysis is used to assess the errors on the fit
parameters: A number of measurements equal to the number of unclipped
data points are randomly selected from the set of unclipped data
points, with replacement after each selection.  These data points are
then used to fit for the astrometric parameters, using ordinary least
squares fitting.  The parameters are recorded and the process re-run
100 times.  For each astrometric parameter, the error is determined as
half of the 68\% confidence range for the distribution of fitted
parameter values.

\section{Discussion}

\section{Conclusion}

\acknowledgments

The Pan-STARRS1 Surveys (PS1) have been made possible through
contributions of the Institute for Astronomy, the University of
Hawaii, the Pan-STARRS Project Office, the Max-Planck Society and its
participating institutes, the Max Planck Institute for Astronomy,
Heidelberg and the Max Planck Institute for Extraterrestrial Physics,
Garching, The Johns Hopkins University, Durham University, the
University of Edinburgh, Queen's University Belfast, the
Harvard-Smithsonian Center for Astrophysics, the Las Cumbres
Observatory Global Telescope Network Incorporated, the National
Central University of Taiwan, the Space Telescope Science Institute,
the National Aeronautics and Space Administration under Grant
No. NNX08AR22G issued through the Planetary Science Division of the
NASA Science Mission Directorate, the National Science Foundation
under Grant No. AST-1238877, the University of Maryland, and Eotvos
Lorand University (ELTE) and the Los Alamos National Laboratory.

\bibliographystyle{apj}

\end{document}